%% file: main.tex
\documentclass{IEEEtran}
\def\BibTeX{{\rm B\kern-.05em{\sc i\kern-.025em b}\kern-.08em
		T\kern-.1667em\lower.7ex\hbox{E}\kern-.125emX}}

\usepackage{cite}
\usepackage{amsmath,amssymb,amsfonts}
\usepackage{mathabx}
\usepackage{algorithm}
\usepackage{algpseudocode}
\usepackage{textcomp}
\usepackage[utf8]{inputenc}
\usepackage[T1]{fontenc}
\usepackage{balance}
\usepackage[rightcaption]{sidecap}
\usepackage{import}
\usepackage{textcomp}
\usepackage[dvipsnames]{xcolor}
\usepackage{slashed}
\usepackage{subcaption}
\usepackage{enumitem}
\usepackage{tabularx}
\usepackage{wrapfig}
\usepackage{bm}
\usepackage{cancel}
\usepackage{caption}
\usepackage{graphicx}
\usepackage{adjustbox}

\usepackage{easyReview}

\newcommand{\bbR}{\mathbb{R}}

\newcommand{\calC}{\mathcal{C}}
\newcommand{\calD}{\mathcal{D}}

\newcommand{\calH}{\mathcal{H}}

\newcommand{\calK}{\mathcal{K}}

\newcommand{\calS}{\mathcal{S}}
\newcommand{\calT}{\mathcal{T}}
\newcommand{\calU}{\mathcal{U}}

\newcommand{\calX}{\mathcal{X}}

\usepackage{amsthm}

\theoremstyle{definition}
\newtheorem{assumption}{Assumption}
\newtheorem{theorem}{Theorem}
\newtheorem{lemma}[theorem]{Lemma}
\newtheorem{proposition}[theorem]{Proposition}
\newtheorem{corollary}[theorem]{Corollary} 
\newtheorem{definition}{Definition}

\newtheorem{example}{Example}

\theoremstyle{remark}
\newtheorem{remark}{Remark}

\begin{document}
\title{On Uniformly Time-Varying Control Barrier Functions}
\author{Adrian Wiltz, and Dimos V. Dimarogonas
\thanks{This work was supported by the ERC Consolidator Grant LEAFHOUND, the Horizon Europe EIC project SymAware (101070802), the Swedish Research Council, and the Knut and Alice Wallenberg Foundation.}
\thanks{The authors are with the Division of Decision and Control Systems, KTH Royal Institute of Technology, SE-100 44 Stockholm, Sweden {\tt\small \{wiltz,dimos\}@kth.se}.}}

\maketitle

\begin{abstract}                          
	\import{}{abstract}
\end{abstract}

\begin{IEEEkeywords}                           
Control Barrier Functions, Constrained Control, Time-Varying Systems, Non-smooth systems, Nonlinear Control
\end{IEEEkeywords}                             


\import{}{00-introduction}


\import{}{01-preliminaries}


\import{}{02-main-results}

\import{}{03-relation-of-lambda-shiftable-CBFs-and-CLFs}

\import{}{05-some-guidelines-on-the-construction-of-alpha-functions}


\import{}{10-simulation}


\import{}{20-conclusion}


\bibliographystyle{IEEEtran}

\balance

\bibliography{/home/adrian-wiltz/MyCloudStation/JabBib/Research/000_MyLibrary}


\end{document}

%% file: abstract.tex
This paper investigates the design of a subclass of time-varying Control Barrier Functions (CBFs), specifically that of uniformly time-varying CBFs. Leveraging the fact that CBFs encode a system’s dynamic capabilities relative to a state constraint, we decouple the design of uniformly time-varying CBFs into a time-invariant and a time-varying component. We characterize the subclass of time-invariant CBFs that yield a uniformly time-varying CBF when combined with a specific type of time-varying function. A detailed analysis of those conditions under which the time-varying function preserves the CBF property of the time-invariant component is provided. These conditions allow for selecting the time-varying function such that diverse variations in the state constraints can be captured while avoiding the redesign of the time-invariant component. From a technical point of view, the analysis requires the derivation of novel relations for comparison functions, not previously reported in the literature. Finally, we discuss how existing CBF construction methods can be applied to design suitable time-invariant CBFs, and demonstrate the effectiveness of the approach through detailed numerical examples.

%% file: 00-introduction.tex
\section{Introduction}

The literature on Control Barrier Functions (CBFs) provides both a broad range of guarantees for constraint satisfaction~\cite{Ames2019,Wabersich2023a,Garg2024a} along with analytical and numerical methods for their construction; a review of these methods is provided later in Section~\ref{sec:cbf and class K function construction}. Most existing works focus on time-invariant constraints with comparatively few works addressing time-variations. Such time-variations may arise in the CBF design, for example, from obstacles whose position or size changes over time~\cite{Tee2011,Sankaranarayanan2024,Haraldsen2024,Namerikawa2024,Lanza2025,Huang2025}, or from control specifications encoded as time-varying state constraints~\cite{Lindemann2019,Yang2020,Wiltz2022a}. Most available design methods for time-varying CBFs, however, require an explicit model of the time-variation of the constraint, or assume the availability of sufficiently large or even unbounded control inputs, see~\cite{Tee2011,Sankaranarayanan2024,Haraldsen2024,Namerikawa2024,Lanza2025,Huang2025,Lindemann2019,Yang2020,Wiltz2022a}. These works address the design of time-varying CBFs predominantly with analytical methods; numerical methods are hindered by the extra dimension introduced through time and the need to consider potentially unbounded time horizons. Even minor deviations from the assumed time-variations require the recomputation of the CBF, resulting in high computational effort. The problem of time-varying constraints or obstacles has been previously also addressed with artificial potential fields~\cite{Ge2002,Loizou2003} or high-gain control approaches~\cite{Berger2021,Mehdifar2025} requiring the dynamics to be input-output linearizable or to possess at least the high-gain controllability property; the incorporation of input constraints into those approaches is challenging if possible at all. In our work, we address dynamics even beyond these limitations as well as systems comprising input constraints.

To the best of our knowledge, the general treatment of time-varying constraints through CBFs remains an open problem. Nevertheless, CBFs -- including specifically the time-invariant case -- provide a rigorous characterization of a system's dynamic capabilities with respect to a state constraint. This characterization can be advantageously exploited to decouple the design of time-varying CBFs into the design of a time-invariant and a time-dependent component.

Motivated by theses considerations, we approach the problem of time-varying constraints by focusing on a particular class of time-varying CBFs, termed \emph{uniformly time-varying CBFs}, taking the form
\begin{align}
	\label{eq:uniformly time-varying CBF}
	B_{\bm{\lambda}(\cdot)}(t,x) \coloneq b(x) + \bm{\lambda}(t),
\end{align}
where $ b:\bbR^{n}\rightarrow\bbR $ depends solely on the state~$ x $ and $ \bm{\lambda}:\bbR_{\geq0}\rightarrow\bbR $ solely on time~$ t $, both to be further specified in subsequent sections. Given a CBF~$ b $, we derive conditions guaranteeing the existence of a function~$ \bm{\lambda} $ such that $ B_{\bm{\lambda}(\cdot)} $ preserves the CBF property of $ b $. Furthermore, we characterize $ \bm{\lambda} $ with respect to its rate of change such that $ B_{\bm{\lambda}(\cdot)} $ remains a valid CBF. Such an approach avoids entangling state- and time-constraints in the design process, enabling the use of established analytical and numerical methods to construct $ b $ independently of the time-variation. The function~$ \bm{\lambda} $ can then be selected to reflect the specific time-variation of the constraint under consideration, without requiring a redesign of $ b $. Thereby, any time-varying state constraint of the form
\begin{align*}
	x\in\calH(t)\subseteq \{x \, | \, h(x) + \bm{\gamma}(t)\geq 0\}
\end{align*}
with $ h:\bbR^{n}\rightarrow\bbR $ and $ \gamma: \bbR_{\geq 0}\rightarrow\bbR $ can be handled.

We point out that time-varying CBFs do not necessarily imply that time-varying constraints are addressed. As such, \cite{Choi2021,Maghenem2022} obtain time-varying barrier functions due to their reachability-based definition on a finite time horizon, while the state constraint under consideration is time-invariant. 

Our contributions are summarized as follows:
\begin{enumerate}
	\item The subclass of time-invariant CBFs $ b $ that admit a uniformly time-varying CBF~$ B_{\bm{\lambda}(\cdot)} $ of the form~\eqref{eq:uniformly time-varying CBF} is characterized. A detailed analysis provides sufficient conditions on the time-varying functions~$ \bm{\lambda} $ that preserve the CBF property of~$ b $ under their summation. Specifically, the extended class~$ \calK_{e} $ function associated with CBF $ b $, commonly denoted by~$ \alpha $, is shown to be directly linked to the rate of change of $ \bm{\lambda} $, emphasizing the need for a sophisticated design of $ \alpha $. 
	\item The construction of an extended class~$ \calK_{e} $ function associated with a uniformly time-varying CBF requires the derivation of novel relations for the sum of a class~$ \calK $ function and an extended class~$ \calK_{e} $ function; to the best of our knowledge, these have not been previously derived, including~\cite{Kellett2014}.
	\item We remark on the design of CBFs belonging to the subclass mentioned in~(1), followed by a detailed study on the application of the proposed framework through various numerical examples. In the latter, particular emphasis is placed on the selection of~$ \alpha $, noting that a linear choice, though prevalent in the literature, is not necessarily the least conservative.  
\end{enumerate}

A preliminary version of the results in contribution~(1) appeared in~\cite{Wiltz2024a}, where both $ b $ and $ \bm{\lambda} $ were assumed to be differentiable. In this paper, these assumptions are relaxed: $ b $ is only required to be a CBF in the Dini sense, and $ \bm{\lambda} $ piecewise differentiable and piecewise continuous. The other contributions are original to this work.

\emph{Notation:} A function $ \alpha: \bbR_{\geq0} \rightarrow \bbR_{\geq0} $ is a \emph{class~$ \calK $ function} if it is strictly increasing and $ \alpha(0) = 0 $. If extended to $ \alpha: \bbR \rightarrow \bbR $, it is called an \emph{extended class~$ \calK_{e} $ function}. Matrices are denoted by uppercase letters, sets by calligraphic letters, and trajectories in boldface. The set of trajectories on $ [t_{1}, t_{2}] $ is $ \bm{\calX}_{[t_{1},t_{2}]} $, abbreviated as $ \bm{\calX} $ when clear from the context. We call a function $f:\bbR\rightarrow\bbR^{n}$ \emph{piecewise differentiable} on $[t_{0},t_{N}]$ if there exists a finite partition $t_{0}<\dots<t_{N}$ such that $f$ is continuously differentiable on each interval $[t_{i},t_{i+1})$. For $ \calS\subseteq\bbR^{n} $, the \emph{Hausdorff-distance} is $ d_{H}(x,\calS) \coloneq \inf_{y\in\calS} ||x-y|| $, and the sign-function is $ \text{sgn}(x) = 1 $ if $ x>0 $, $ 0 $ if $ x=0 $, and $ -1 $ otherwise. For $ x\in\bbR^{n} $, $ Q\in\bbR^{n\times n} $, and $ p\geq1 $, $ ||x||_{p} $ denotes the $ p $-norm, $ ||Q||_{p} $ its induced matrix norm, and $ ||x||_{Q}^{2}\coloneq x^{T} Q x $. The $ n $-dimensional identity and zero matrices are denoted by $ \bm{I}_{n} $ and $ \bm{0}_{n} $, respectively; when clear from the context, we write $ \bm{I} $ and $ \bm{0} $, where $ \bm{0} $ may also have rectangular shape when appropriate.

%% file: 01-preliminaries.tex
\section{Preliminary Results}
\label{sec:preliminaries}

We begin by reviewing CBFs and CBFs in the Dini sense. We then introduce time-varying CBFs in the Dini sense, relate them to existing notions, and establish a forward invariance result.

Let us consider a dynamical system subject to input constraints, given as
\begin{align}
	\label{eq:dynamics}
	\dot{x} = f(x,u), \qquad x(0) = x_{0},
\end{align}
where $ x, x_{0}\in\bbR^{n} $, $ u\in\calU\subseteq\bbR^{m} $, and $ f: \bbR^{n}\times\calU \rightarrow\bbR^{n} $ is Lipschitz continuous in both arguments to ensure the uniqueness of its solutions. It is assumed that the system is forward complete. The state trajectory corresponding to a piecewise continuous input trajectory $ \bm{u}: \bbR_{\geq0} \rightarrow \bm{\calU} $ is denoted by $ \bm{\varphi}(t;x_{0},\bm{u}) $, where $ t $ is the time at which $ \bm{\varphi} $ is evaluated. A CBF is now formally defined as follows.

\begin{definition}\textbf{\emph{(CBF~\cite{Ames2017},\cite{Wieland2007})}}
	\label{def:cbf}
	Consider $ \calD\subseteq\bbR^{n} $ and a continuously differentiable function $ b: \bbR^{n}\rightarrow\bbR $ such that the zero super-level set of $ b $ defined as   
	\begin{align}
		\label{eq:calC}
		\calC := \{ x \, | \, b(x)\geq 0 \}
	\end{align}
	is closed and it holds $ \calC\subseteq\calD\subseteq\bbR^{n} $. We call such $ b $ a \emph{Control Barrier Function} (CBF) on $ \calD $ with respect to~\eqref{eq:dynamics} if there exists an extended class $ \calK_{e} $ function $ \alpha $ such that for all $ x\in\calD $ 
	\begin{align}
		\label{eq:def cbf 1}
		\sup_{u\in\calU} \left\{ \frac{\partial b}{\partial x}(x) \, f(x,u) \right\} \geq -\alpha (b(x)). 
	\end{align}
\end{definition}

Let us now consider the (time-invariant) state constraint
\begin{align}
	\label{eq:time-invarinat state constraint}
	x(t) \in \calH := \{x \, | \, h(x)\geq 0\} \qquad \forall t\geq 0,
\end{align}
where $ h: \bbR^{n}\rightarrow\bbR $ is a Lipschitz continuous function. A CBF $ b $ can be viewed as a system theoretic characterization of the dynamical capabilities of a system~\eqref{eq:dynamics} with respect to a constraint~\eqref{eq:time-invarinat state constraint}, when $ b(x) \leq h(x) $ for all $ x\in\calC $. Then, $ \calC\subseteq\calH $.
Set $ \calC $ is called \emph{forward control invariant} with respect to system~\eqref{eq:dynamics} if there exists an input trajectory $ \bm{u}\in\bm{\calU}_{[0,\infty)} $ such that $\bm{\varphi}(t;x_{0},\bm{u})\in\calC$ for all $ t\geq 0 $. Furthermore, $ \calC $ is called \emph{forward invariant} under a given input $ \bm{u}\in\bm{\calU}_{[0,\infty)} $ with respect to system~\eqref{eq:dynamics} if $\bm{\varphi}(t;x_{0},\bm{u}) $ remains in $ \calC$ for all $ t\geq 0 $. 

Generally, however, it is limiting to require CBFs to be continuously differentiable functions. The reasons for this are twofold. Firstly, if constraint $ h $ is only Lipschitz continuous instead of differentiable everywhere, then a continuously differentiable CBF may only render a strict subset $ \calC \subset \calH $ forward invariant. For instance, this is the case for the $ n $-dimensional single-integrator $ \dot{x} = u $ subject to a constraint defined as a hyper-cube~$ \calH $ given through $ h(x) = 1-\max_{i} \{x_{i}\} \geq 0 $. In this example, $ \calC $ must be an inner-approximation of $ \calH $ for the sake of smoothness of the defining CBF, whereas clearly also $ \calH $ could be rendered forward invariant. Secondly, as CBFs are closely related to Control Lyapunov Functions (CLF), the need for a notion with relaxed smoothness properties is inherited~\cite{Sontag1983,Kellett2004}; we illustrate this with numerical examples in Sections~\ref{subsec:example nonholonomic system} and~\ref{subsec:example time-varying obstacle}, where input constrained and non-holonomic systems are considered. Motivated by this, we introduce \emph{CBFs in the Dini sense} analogously to CLFs in the Dini sense~\cite{Sontag1983,Clarke2011}.

\begin{definition}[CBF in the Dini Sense \cite{Wiltz2025b}]
	\label{def:cbf dini}
	Consider $ \calD\subseteq\bbR^{n} $ and a locally Lipschitz continuous function $ b: \bbR^{n} \rightarrow \bbR $ such that $ \calC $ as defined in~\eqref{eq:calC} is closed and $ \calC\subseteq\calD\subseteq\bbR^{n} $. We call such $ b $ a \emph{CBF in the Dini sense} on $ \calD $ with respect to~\eqref{eq:dynamics} if there exists an extended class~$ \calK_{e} $ function $ \alpha $ such that for all~$ x\in\calD $
	\begin{align}
		\label{eq:def cbf dini}
		\sup_{u\in\calU} \left\{ db(x;f(x,u)) \right\} \geq -\alpha(b(x)),
	\end{align}
	where $ db(x;v) $ denotes the \emph{Dini derivative} at $ x $ in direction~$ v $ as
	\begin{align}
		\label{eq:dini}
		db(x;v) \coloneq \lim_{\varepsilon \downarrow 0}\frac{b(x+\varepsilon v) - b(x)}{\varepsilon};
	\end{align}
	we assume that the upper and lower Dini derivatives coincide, i.e., $d^{+}b(\cdot;\cdot) \!\!\equiv\!\! d^{-}\!b(\cdot;\cdot) $ with $d^{+}b(x;v) \!\!\coloneq\!\! \limsup_{\varepsilon \downarrow 0}\!\frac{b(x\!+\!\varepsilon v) \!-\! b(x)}{\varepsilon}$ and $d^{-}\!b(x;v) \!\!\coloneq\!\! \liminf_{\varepsilon \downarrow 0}\!\frac{b(x\!+\!\varepsilon v) \!-\! b(x)}{\varepsilon}$.
\end{definition}
\begin{remark}
	If $ b $ is continuously differentiable in the neighborhood of $ x $, then $ db(x;f(x,u)) = \frac{\partial b}{\partial x}(x) \, f(x,u) $. Thus, for $ b $ continuously differentiable everywhere, Definitions~\ref{def:cbf} and~\ref{def:cbf dini} coincide. Assuming that the upper and lower Dini derivatives coincide excludes some pathologic, practically less relevant cases, simplifying the exposition.
\end{remark}

Moreover, we introduce an explicitly time-varying version of the latter definition with relaxed continuity properties on the time argument. In particular, we assume that the time-varying function $ b: \bbR_{\geq 0} \times \bbR^{n} \rightarrow \bbR $, given at time~$ t $ and state~$ x $ as $ b(t,x) $, satisfies the following relaxed continuity properties.

\begin{assumption}[Relaxed Continuity]
	\label{ass:continuity prop b}
	Function $ b $ is locally Lipschitz continuous with respect to its state argument, piecewise differentiable with respect to its time argument, and it holds
	\begin{align}
		\label{eq:ass continuity prop b}
		\lim_{t\uparrow t_{0}} b(t,x) \leq \lim_{t\downarrow t_{0}} b(t,x) \qquad\forall x\in\bbR^{n}, \; \forall t_{0}\geq 0.
	\end{align} 
	Specifically at those $ (t,x) $, where $ b $ is discontinuous, inequality~\eqref{eq:ass continuity prop b} is strict.
\end{assumption} 

\begin{definition}\textbf{\emph{(Time-Varying CBF in the Dini Sense)}}
	\label{def:cbf dini time-varying}	
	Consider $ \calD \subseteq \bbR^{n} $ and a function $ b: \bbR_{\geq 0} \times \bbR^{n} \rightarrow \bbR $ satisfying Assumption~\ref{ass:continuity prop b} such that $ \calC(t) := \{x \, | \, b(t,x) \geq 0\} $ is closed and it holds $ \calC(t) \subseteq \calD \subseteq \bbR^{n} $ for all $ t\geq 0 $. We call such $ b $ a \emph{time-varying CBF in the Dini Sense} on~$ \calD $ with respect to~\eqref{eq:dynamics} if there exists an extended class~$ \calK_{e} $ function $ \alpha $ such that for all $ (t,x)\in\bbR_{\geq0} \times \calD $ 
	\begin{align}
		\label{eq:def cbf dini time-varying}
		\sup_{u\in\calU} \left\{ db(t,x;1,f(x,u)) \right\} \geq -\alpha(b(t,x)),
	\end{align}
	where $ db(t,x;1,v) := \lim_{\varepsilon \downarrow 0} \frac{b(t+\varepsilon,x+\varepsilon v) - b(t,x)}{\varepsilon} $ is the Dini derivative analogously to before; as previously, the upper and lower Dini derivatives are assumed to coincide.
\end{definition}
\begin{remark}
	\label{rem:eq:ass continuity prop b}
	Property~\eqref{eq:ass continuity prop b} ensures that the zero super-level set of $ b $ satisfies at discontinuities 
	\begin{align*}
		\lim_{t\uparrow t_{0}} \calC(t) \subset \lim_{t\downarrow t_{0}} \calC(t).
	\end{align*}
	Thus, any state $ x $ contained before the discontinuity in $ \calC(t) $ is also contained afterwards. Assumption~\ref{ass:continuity prop b} allows for generalized time-variations beyond differentiable ones, and is to be ensured during the construction of~$ b $. We relate it later to the construction of uniformly time-varying CBFs.
\end{remark}

The latter definition includes Definitions~\ref{def:cbf} and~\ref{def:cbf dini} as special cases. 
The forward invariance result for CBFs (also for those in the time-varying and the Dini sense) is a corollary of Nagumo's theorem.

\begin{theorem}\emph{\textbf{(Nagumo's Theorem)}}
	\label{thm:Nagumo}
	Consider dynamic system~\eqref{eq:dynamics} and let $ \bm{u}: \bbR_{\geq0} \rightarrow \calU $ be continuous in time. Let $ \calS\subseteq\bbR^{n} $ be a closed set. Moreover, we define the tangent cone to $ \calS $ at $ x $ as $ \calT_{\calS}(x) := \big\lbrace v \, \big| \, \liminf_{\varepsilon\rightarrow 0} \frac{d_{H}(x+\varepsilon v,\calS)}{\varepsilon} = 0 \big\rbrace $, where $ d_{H} $ denotes the Hausdorff distance. Then, $ \calS $ is forward control invariant with respect to~\eqref{eq:dynamics} if and only if, for every $ x_{0}\in\calS $, there exists some $ \bm{u} $ such that 
	\begin{align}
		\label{eq:thm:Nagumo condition}
		f(x_{t},u_{t}) \in \calT_{\calS}(x_{t}) \qquad \forall t\in[0,\varepsilon]
	\end{align}
	for some $\varepsilon>0$, where $ x_{t} \coloneq \bm{\varphi}(t;x_{0},\bm{u}) $ and $ u_{t} \coloneq \bm{u}(t) $.
\end{theorem}
\begin{remark}
	The presented theorem is a direct corollary of the actual theorem by Nagumo (cf.~\cite[Theorem~4.7 and Corollary~4.8]{BlanchiniFranco2015SMiC}). We note that in the context of a time-varying CBF~$ b(t,x) $ in the Dini sense, \eqref{eq:thm:Nagumo condition} becomes $ \left[\begin{smallmatrix} 1 \\ f(x,u) \end{smallmatrix}\right] \in \calT_{\calS}(x) $ and~$ \calS \!=\! \{ (t,x) \, | \, (t,x) \!\in\! \{t\}\!\times\!\calC(t), \; t \!\geq\!0\} $. Furthermore, as $ \calC $, and thereby $ \calS $, are characterized through $ b $, condition~\eqref{eq:thm:Nagumo condition} reduces to $ db(t,\bm{x}(t);1,f(\bm{x}(t),\bm{u}(t))) \geq 0 $ for $ x_{0} $ on the boundary of~$ \calS $. Due to~\eqref{eq:ass continuity prop b} and the fact stated in Remark~\ref{rem:eq:ass continuity prop b}, this even holds for times~$ t $, where $ b $ is discontinuous.
\end{remark}

The continuous control input trajectory required by Nagumo's theorem can be obtained through a continuous state feedback controller. In the context of CBFs, \cite{Wieland2007,Ames2017} provides a constructive method for obtaining such feedback controllers for input affine systems based on differentiable CBFs. For CBFs in the Dini sense, however, the continuity requirement needs to be relaxed. Instead of requiring the continuity of the state feedback controller, we now only ask for a feedback controller that does not exhibit Zeno behavior~\cite{Zhang2001a}; that is, given a state feedback controller $ k: \bbR^{n}\rightarrow \calU $ and $ k\circ x $ with $ x $ governed by~\eqref{eq:dynamics} is discontinuous at $ t $, then there may be no further discontinuouties in an arbitrarily small $ \varepsilon $-neighborhood of $ t $. The resulting input trajectory $ \bm{u} $ is thereby piecewise continuous. For CBFs in the Dini sense and input affine dynamics, such feedback controller is derived in~\cite{Wiltz2022a} based on Dini-derivatives. These considerations lead to a corollary of Nagumo's theorem deriving the forward invariance of $ \calC(t) $ via time-varying CBFs in the Dini-sense.

\begin{corollary}[Forward Invariance]
	\label{corollary:cbf invariance}
	Let $ b $ be a time-varying CBF in the Dini sense with respect to~\eqref{eq:dynamics} on $ \calD\subseteq\bbR^{n} $. Furthermore, let $ \bm{u}: \bbR_{\geq0} \rightarrow \calU $ be piecewise continuous and the corresponding state trajectory $ \bm{\varphi}(\cdot;x_{0},\bm{u}) $ starting in some initial state $ x_{0}\in\calC(0) $. If 
	\begin{align}
		\label{eq:corollay:cbf invariance}
		db(t,x_{t};1,f(x_{t},u_{t})) \geq -\alpha(b(t,x_{t}))	\qquad\!\!\! \forall t\in[0,T],
	\end{align} 
	where $ x_{t} \coloneq \bm{\varphi}(t;x_{0},\bm{u}) $ and $ u_{t}\coloneq \bm{u}(t) $, and $ \alpha $ being the same as in Definition~\ref{def:cbf dini time-varying}, then $ \calC(\cdot) $ is forward invariant such that $ x_{t}\in\calC(t) $ for all $ t\in[0,T] $.
\end{corollary}
\vspace{-\baselineskip}
\begin{proof}
	Without loss of generality, we denote those intervals, on which $ \bm{u} $ is continuous, as $ [\tau_i, \tau_{i+1}) $, where it holds for $ \{\tau_{i}\}_{i=0,\dots,N} $ that
	\begin{align*}
		\tau_{0} = 0 < \cdots < \tau_{i} < \tau_{i+1} < \cdots < \tau_{N} = T. 
	\end{align*}
	Now, let $ t\in[\tau_{i},\tau_{i+1}) $. Note that \eqref{eq:corollay:cbf invariance} implies that $ f(x_{t},u_{t}) $ is contained in the tangent cone $ \calT_{\calS}(x_{t}) $ of $ \calS \coloneq \{ (t,x) \, | \, (t,x) \in \{t\}\times\calC(t), \; t\geq 0\} $ and $ \bm{u} $ is continuous on the interval under consideration. Here, $ \calS $ is the time-augmented version of $ \calC(\cdot) $. If $ (\tau_{i},x_{\tau_{i}})\in\calS $, or equivalently $ x_{\tau_{i}}\in\calC(\tau_{i}) $, then the forward invariance of $ \calS $ and $ \calC(\cdot) $ follows on the interval $ [\tau_{i},\tau_{i+1}) $. Together with the continuity properties of $ b $ in Assumption~\ref{ass:continuity prop b}, it follows 
	\begin{align*}
		b(x_{\tau_{i+1}}) \geq \lim_{\tau \uparrow \tau_{i+1}} b(\bm{\varphi}(\tau;x_{\tau_{i}},u)) \geq 0
	\end{align*}
	and $ x_{\tau_{i+1}}\in\calC(\tau_{i+1}) $. As $ x_{0}\in\calC(0) $ and thus $ x_{0}\in\calS $, it follows inductively that $ \calS $ is forward invariant and $ x_{t}\in\calC(t) $ for all $ t\in[0,T] $.
\end{proof}

%% file: 02-main-results.tex
\section{Uniformly Time-Varying CBFs}
\label{sec:main results}

We now derive our main results on uniformly time-varying CBFs by first introducing a particular class of CBFs, termed shiftable CBFs, and then deriving our main theorem on their compositional construction, which decomposes the design into a purely state-dependent and a purely time-dependent component.

The construction of time-varying CBFs is generally demanding as it requires to determine a CBF over the time-augmented state space. This is specifically demanding when constructing a CBF beyond a finite time horizon. For time-invariant CBFs, in contrast, there exist a multitude of construction methods. Same holds for the closely related Control Lyapunov Functions (CLFs). We aim in this section at leveraging the information on the system's dynamic properties encoded in these function to derive uniformly time-varying CBFs of the form
\begin{align*}
	b(x) + \bm{\lambda}(t),
\end{align*}
where $ b $ is a particular type of CBF, which we call \emph{shiftable}, and $ \bm{\lambda} $ is a time-varying function. 

\subsection{Shiftable CBFs}

We first define the class of \emph{shiftable CBFs} and their extension in the Dini sense, which constitute a subclass of differentiable CBFs and CBFs in the Dini sense, respectively.

\begin{definition}[$ \Lambda $-shiftable CBF]
	\label{def:lambda shiftable CBF}
	A Lipschitz continuous function $ b:\bbR^{n}\rightarrow\bbR_{\geq 0} $ is called a \emph{$ \Lambda $-shiftable CBF} \emph{(in the Dini sense)} with respect to~\eqref{eq:dynamics} for some $ \Lambda > 0 $ if $ b $ is a CBF (in the Dini sense) on the domain 
	\begin{align*}
		\calC_{\Lambda} := \{ x \, | \, b(x) \geq -\Lambda \}
	\end{align*}
	with respect to~\eqref{eq:dynamics}, or equivalently, if there exists an extended class~$ \calK_{e} $ function~$ \alpha $ such that~\eqref{eq:def cbf 1} (respectively \eqref{eq:def cbf dini}) holds for all $ x\in\calC_{\Lambda} $.
\end{definition}
\begin{remark}
	For the sake of a more compact terminology, we adopt from now on the following naming convention: whenever we write ($ \Lambda $-shiftable or time-varying) CBF in the sequel, we implicitly refer to their generalization in the Dini sense unless stated otherwise. 
\end{remark}

In the previous definition, $ \calC_{\Lambda} $ takes the role of domain~$ \calD $ in Definitions~\ref{def:cbf} and~\ref{def:cbf dini}. We call such CBF $ \Lambda $-shiftable since $ b(x)\!+\!\lambda $, which is the by $ \lambda $ shifted version of $ b(x) $, is still a CBF for any $ \lambda\!\in\![0,\Lambda] $. 

\begin{proposition}
	\label{prop:lambda_constant_shifted}
	Let $ \lambda\in[0,\Lambda] $ and $ b $ be a $ \Lambda $-shiftable CBF. Then the by $ \lambda $ shifted function
	\begin{align}
		\label{eq:prop:B_lambda}
		B_{\lambda}(x) := b(x) + \lambda
	\end{align}
	is a CBF on $ \calC_{\Lambda} $. Moreover, if $ \lambda < \Lambda $, then $ B_{\lambda} $ is also a $ (\Lambda\!-\!\lambda) $-shiftable CBF. 
\end{proposition}

\begin{proof}
	For the first part, we derive that 
	\begin{align*}
		&\sup_{u\in\calU} \left\{ dB_{\lambda}(x;f(x,u)) \right\} \stackrel{\eqref{eq:dini}}{=} \sup_{u\in\calU} \left\{ db(x;f(x,u)) \right\} \\
		&\qquad \stackrel{\text{\eqref{eq:def cbf dini}}}{\geq} -\alpha(b(x)) \geq -\alpha(b(x)+\lambda) = -\alpha(B_{\lambda}(x)) 
	\end{align*}
	for all $ x\in\calC_{\Lambda} $. Thus by Definition~\ref{def:cbf dini}, $ B_{\lambda} $ is a CBF on $ \calC_{\Lambda} $. Since $ \calC_{\Lambda} = \{ x \, | \, b(x) \geq -\Lambda \} = \{ x \, | \, B_{\lambda}(x) \geq -(\Lambda-\lambda) \} $, $ B_{\lambda} $ is a $ (\Lambda\!-\!\lambda) $-shiftable CBF if $ \Lambda - \lambda > 0 $, or equivalently,  $ \lambda < \Lambda$, and the second part follows as well.
\end{proof}

Clearly, every $ \Lambda $-shiftable CBF is a CBF, but not every CBF is $ \Lambda $-shiftable. To illustrate this, let us consider the exemplary system
\begin{align*}
	\dot{x}_{1} &= (x_{1}^2+x_{2}^2-u)x_{2}, \\
	\dot{x}_{2} &= (x_{1}^2+x_{2}^2-u)x_{1},
\end{align*}
which is subject to the input constraint $ |u|\leq1 $, and a function 
\begin{align*}
	b(x_{1},x_{2}) = -x_{1}^2 - x_{2}^2 + 1,
\end{align*}
where $ \calC = \{ [x_{1},x_{2}]^{T}\in\bbR^2 \, | \, b(x_{1},x_{2})\geq 0\} = \{ [x_{1},x_{2}]^{T}\in\bbR^2 \, | \, ||[x_{1},x_{2}]||^{2}\leq 1\} $ is the unit circle.
Function $ b $ is a CBF in the sense of Definition~\ref{def:cbf dini} since for all $ (x_{1},x_{2})\in\calC $ and some $ u $ with $ |u|\leq1 $, for example $u=1$, it holds
\begin{align*}
	\frac{db}{dt}(x_{1},x_{2}) \!&=\! -4x_{1}x_{2}(x_{1}^2\!+\!x_{2}^2\!-\!u) \!=\! 4x_{1}x_{2} (b(x_{1},x_{2})\!-\!1\!+\!u) \\
	\!&\geq\! -4 b(x_{1},x_{2}) \!=\! -\alpha(b(x_{1},x_{2})),
\end{align*}
where $ \alpha(b) \coloneq 4b $ is a class $ \calK $ function. However, computing the Lie-derivative of $ B_{\Lambda}(x_{1},x_{2}) = b(x_{1},x_{2}) + \Lambda $, with $ \Lambda > 0 $, reveals that for certain points $ (x_{1},x_{2})\!\in\!\calC_{\Lambda} = \{ [x_{1},x_{2}]^{T}\, | \; ||[x_{1},x_{2}]||^{2}\!\leq\! 1\!+\!\Lambda \} $ (specifically those with $ x_{1}x_{2}>0 $) and all inputs $ u $ with $ |u|\leq1 $, we have
\begin{align*}
	\frac{dB_{\Lambda}}{dt}(x_{1},x_{2}) &= -4x_{1}x_{2}(x_{1}^2\!+\!x_{2}^2\!-\!u) \\
	&\hspace{-1.1cm}\stackrel{||[x_{1},x_{2}]||^{2}\leq 1\!+\!\Lambda}{\leq}\!\!\! -4x_{1}x_{2} (1\!+\!\Lambda\!-\!u) \hspace{-0.15cm}\stackrel{x_{1}x_{2}>0}{\leq}\!\! -4x_{1}x_{2}\Lambda < 0
\end{align*}
for all $ \Lambda > 0 $. Thus, \eqref{eq:def cbf 1} is only satisfied on $ \calC = \calC_{\Lambda}\big|_{\Lambda = 0}  $, and $ b(x_{1},x_{2}) $ is not $ \Lambda $-shiftable. 

By shifting a CBF, its zero super-level set can be uniformly varied. As such, a shift of a $ \Lambda $-shiftable CBF by $ \lambda\in[0,\Lambda] $ leads to the invariance of $ \calC_{\lambda} $ via Corollary~\ref{corollary:cbf invariance}.

\subsection{Constructing Uniformly Time-Varying CBFs from Shiftable CBFs}

We now transition from a constant shift $ \lambda $ to a time-dependent function $ \bm{\lambda}(\cdot) $. It is assumed to satisfy the following continuity properties. 

\begin{assumption}
	\label{ass:lambda prop}
	 Function $ \bm{\lambda}(\cdot): \bbR_{\geq 0} \rightarrow [0,\Lambda]\subseteq\bbR_{\geq0} $ is piecewise differentiable, and it holds
	\begin{align*}
		\lim_{t\uparrow t_{0}} \bm{\lambda}(t) < \lim_{t\downarrow t_{0}} \bm{\lambda}(t) \qquad\forall t_{0}\in\Omega,
	\end{align*}
	where $ \Omega := \{ t \, | \, \bm{\lambda}(\cdot) \text{ is discontinuous at time } t \} $.
\end{assumption}

\begin{figure}[t]
	\centering
	\def\svgwidth{0.5\columnwidth}
	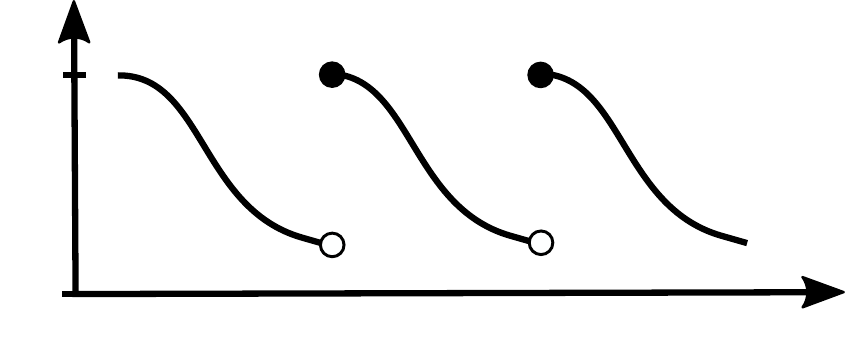
	\caption{Example of $ \bm{\lambda}(\cdot) $ with continuity properties as required by Assumption~\ref{ass:lambda prop}.}
	\label{fig:lambda_function_sketch}
\end{figure}

This is analogous to Assumption~\ref{ass:continuity prop b}. An exemplary function with the specified continuity properties is depicted in Figure~\ref{fig:lambda_function_sketch}. By adding $ \bm{\lambda}(\cdot) $ to a $ \Lambda $-shiftable CBF, we obtain a time-varying value function of the form
\begin{align}
	\label{eq:B_lambda_time-varying}
	B_{\bm{\lambda}(\cdot)}(t,x) := b(x) + \bm{\lambda}(t).
\end{align}
The CBFs considered in \cite{Lindemann2019,Wiltz2022a} take this form. 

Our objective is to design the time-varying function~$ \bm{\lambda}(\cdot) $ for a given $ \Lambda $-shiftable CBF $ b $, such that $ B_{\bm{\lambda}(\cdot)}(t,x) $ constitutes a time-varying CBF with respect to dynamics~\eqref{eq:dynamics}. This allows us, together with Corollary~\ref{corollary:cbf invariance}, to guarantee the forward invariance of the uniformly time-varying set
\begin{align}
	\label{eq:C_lambda}
	\calC_{\bm{\lambda}(\cdot)}(t):=\{ x \, | \, b(x) \geq -\bm{\lambda}(t) \}.
\end{align}

To this end, we require that $ \bm{\lambda}(t) $ satisfies, for all $ t\geq 0 $,
\begin{align}
	\label{eq:lambda_dot condition}
	d\bm{\lambda}(t;1) \geq -\alpha_{\lambda}(\bm{\lambda}(t)),
\end{align}
where $ \alpha_{\lambda} $ is a class $ \calK $ function yet to be further specified. Inequality~\eqref{eq:lambda_dot condition} imposes a lower-bound on the rate of change of~$ \bm{\lambda} $. Intuitively, the condition implies that as $ \bm{\lambda} $ approaches the lower bound of its range, which is zero by convention, its rate tends to zero accordingly. Under this assumption, we can establish that $ B_{\bm{\lambda}(\cdot)}(t,x) $ is a time-varying CBF; the result is proven in the next section.

\begin{theorem}[Uniformly Time-Varying CBF]
	\label{thm:time-varying CBF without input constraints}
	Let $ B_{\bm{\lambda}(\cdot)}(t,x) := b(x) + \bm{\lambda}(t) $, where $ b $ is a $ \Lambda $-shiftable CBF with respect to~\eqref{eq:dynamics} with input constraint $ u\in\calU\subseteq\bbR^{m} $, and $ \alpha $ the extended class~$ \calK_{e} $ function associated with~$ b $ as by Definition~\ref{def:lambda shiftable CBF}. Moreover, let $ \bm{\lambda}: \bbR_{\geq0}\rightarrow[0,\Lambda] $ satisfy Assumption~\ref{ass:lambda prop} and condition~\eqref{eq:lambda_dot condition}, where $ \alpha_{\lambda} $ is a convex or concave class $ \calK $ function and $ \alpha(-\xi)\leq -\alpha_{\lambda}(\xi) $ for all $ \xi\in[0,\Lambda] $. Then, $ B_{\bm{\lambda}(\cdot)}(t,x) $ is a time-varying CBF on $ \calC_{\Lambda} $ with respect to dynamics~\eqref{eq:dynamics}.
	Notably, then there exists an extended class~$ \calK_{e} $ function $ \beta $ such that for all $ (t,x)\in\bbR_{\geq 0} \times \calC_{\Lambda} $
	\begin{align}
		\label{eq:thm:time-varying CBF}
		\sup_{u\in\calU} \left\{ dB_{\bm{\lambda}(\cdot)} (t,x; 1, f(x,u) ) \right\} \geq -\beta(b(x)+\bm{\lambda}(t)).
	\end{align}
\end{theorem}

\begin{remark}
	The condition that $ \alpha(-\xi)\leq -\alpha_{\lambda}(\xi) $ for all $ \xi\in[0,\Lambda] $ allows for an intuitive interpretation. The function $ \alpha $ on the left-hand side characterizes the minimal possible ascend rate of the system state on the $ \Lambda $-shiftable CBF~$ b $, while $ \alpha_{\lambda} $ on the right-hand side is bounding the rate with which~$ b $ is shifted. Consequently, the condition ensures that function $ b $ is not shifted faster than the system state can move towards the interior of $ \calC_{\bm{\lambda}(\cdot)} $. For constructing $ \beta $ explicitly in terms of $ \alpha $ and $ \alpha_{\lambda} $, we characterize the monotonicity property of $ \alpha_{\lambda} $ in terms of convexity and concavity.
\end{remark}

If the extended class~$ \calK_{e} $ function $ \alpha $ associated with the $ \Lambda $-shiftable CBF $ b $ is readily convex or concave on $ \bbR_{\leq0} $, the previous theorem can be formulated more directly as follows. 

\begin{corollary}
	\label{corollary:time-varying CBF}
	Let $ B_{\bm{\lambda}(\cdot)} $ be as defined in Theorem~\ref{thm:time-varying CBF without input constraints}. If $ \bm{\lambda}:\bbR_{\geq 0} \rightarrow [0,\Lambda] $ satisfies Assumption~\ref{ass:lambda prop} and 
	\begin{align}
		\label{eq:corollary:time-varying CBF}
		d\bm{\lambda}(t;1) \geq \alpha(-\bm{\lambda}(t)) \qquad \forall t\geq 0,
	\end{align}
	where $ \alpha $ is an extended class~$ \calK_{e} $ function convex or concave on~$ \bbR_{\leq0} $ and associated with $ b $, then $ B_{\bm{\lambda}(\cdot)} $ is a time-varying CBF on $ \calC_{\Lambda} $ with respect to dynamics~\eqref{eq:dynamics} and there exists an extend class~$ \calK_{e} $ function $ \beta $ such that~\eqref{eq:thm:time-varying CBF} holds for all $ (t,x)\in\bbR_{\geq 0} \times \calC_{\Lambda} $.
\end{corollary}

\subsection{Proof of Theorem~\ref{thm:time-varying CBF without input constraints}}

The proof of Theorem~\ref{thm:time-varying CBF without input constraints} relies on two intermediate results. In particular, we need to determine under which conditions there exists an extended class~$ \calK_{e} $ function~$ \beta $ that upper-bounds the sum of~$ \alpha $ and~$ \alpha_{\lambda} $ such that
\begin{align}
	\label{eq:class K function sum is upper-bounded}
	\alpha(x_{1}) + \alpha_{\lambda}(x_{2}) \leq \beta(x_{1}+x_{2}).
\end{align}
Clearly, this holds with equality if $ \alpha $ and $ \alpha_{\lambda} $ are linear. However, when such strong assumptions do not hold, more sophisticated conditions are required. These are derived in the subsequent lemmas.

\begin{lemma}
	\label{lemma:time-varying CBF without input constraints 1}
	Let $ \alpha_{1}: \bbR \rightarrow \bbR $ be an extended class $ \calK_{e} $ function, and $ \alpha_{2}: \bbR_{\geq 0} \rightarrow \bbR $ a \emph{convex} class $ \calK $ function such that $ \alpha_{1}(-x)\leq -\alpha_{2}(x) $ for all $ x\in[0,A] $ and some finite $ A>0 $. Then, there exists an extended class~$ \calK_{e} $ function $ \beta $ such that for all $ x_{1} \in [-A,\infty) $, $ x_{2} \in [0,A] $, it holds
	\begin{align}
		\label{eq:time-varying CBF without input constraints 1}
		\alpha_{1}(x_{1}) + \alpha_{2}(x_{2}) \leq \beta(x_{1}+x_{2}).
	\end{align}
\end{lemma}
\begin{proof}
	Before we start with the actual proof, we recall some properties of convex functions that the proof is based on. At first, recall that for a convex function $ \alpha' $ it holds for all $ x,y\in\bbR $ and $ \sigma\in[0,1] $ that
	\begin{align}
		\label{eq:time-varying CBF without input constraints 1 aux 00}
		\alpha'(\sigma x + (1\!-\!\sigma)y) \leq \sigma \alpha'(x) + (1\!-\!\sigma) \alpha'(y).
	\end{align}
	If $ y=0 $ and $ \alpha'(0)=0 $ (e.g., if $ \alpha' $ is a convex class~$ \calK $ function), this implies $ \alpha'(\sigma x) \leq \sigma \alpha'(x) $ for $ \sigma\in[0,1] $. Moreover, when $ \alpha'(0)=0 $, then we also have that
	$ \alpha' $ is superadditive for positive real numbers; that is, for all $ x,y\geq 0 $, it holds
	\begin{align}
		\label{eq:time-varying CBF without input constraints 1 aux 0}
		\alpha'(x)+\alpha'(y) \leq \alpha'(x+y).
	\end{align}
	This can be shown as
	\begin{align*}
		\alpha'(x) \!+\! \alpha'(y) 
		&= \alpha'\!\left( \!(x\!+\!y) \frac{x}{x\!+\!y} \right) + \alpha'\!\left( \!(x\!+\!y) \frac{y}{x\!+\!y} \right) \\
		&\hspace{-0.5cm}\leq \frac{x}{x\!+\!y}\, \alpha'(x\!+\!y) + \frac{y}{x\!+\!y}\, \alpha'(x\!+\!y) = \alpha'(x\!+\!y),
	\end{align*}
	where the inequality results from the fact that $ \alpha'(\sigma x) \leq \sigma \alpha'(x) $ for $ \sigma\in[0,1] $ as previously observed.
	
	Furthermore, we recall that the difference quotient $ D_{\alpha'}(x,y) := \frac{\alpha'(y)-\alpha'(x)}{y-x} $ of the convex function $ \alpha' $, where $ x < y $, is monotonously increasing in both of its arguments. This is a standard result and it can be easily shown as follows. At first, let $ x $ be fixed, and choose $ y = \sigma y' + (1-\sigma)x  $ where $ y'>x $; thus, $ y \leq y' $. Then, it follows that $ D(x,y) $ is monotonously increasing in~$ y $ as 
	\begin{align*}
		D(x,y) &= \frac{\alpha'(y)-\alpha'(x)}{y-x} = \frac{\alpha'(\sigma y' + (1-\sigma)x) - \alpha'(x)}{\sigma y' + (1-\sigma)x - x}  \\
		&= \frac{\alpha'(\sigma y' + (1-\sigma)x) - \alpha'(x)}{\sigma y' - \sigma x}\\
		&\stackrel{\eqref{eq:time-varying CBF without input constraints 1 aux 00}}{\leq} \frac{(1-\sigma)\alpha'(x) + \sigma \alpha'(y') - \alpha'(x)}{\sigma y' - \sigma x} \\
		&= \frac{\alpha'(y')-\alpha'(x)}{y'-x} = D(x,y').
	\end{align*}
	The result follows for the first argument analogously. 
	Thus, we have $ \frac{\alpha'(y)-\alpha'(x)}{y-x} \leq \frac{\alpha'(y+c)-\alpha'(x+c)}{y-x} $ for all $ x<y $, $ c\geq 0 $, or equivalently,
	\begin{align}
		\label{eq:time-varying CBF without input constraints 1 aux 1}
		\alpha'(y)-\alpha'(x) \leq \alpha'(y+c)-\alpha'(x+c).
	\end{align}
	At last, in addition to its convexity, we assume that $ \alpha' $ is an extended class~$ \calK_{e} $ function. Then, it holds for all $ \sigma \in [0,1/2] $, $ x\geq 0 $, that $ 0 \leq \alpha'((1-2\sigma)x) \leq \sigma \alpha'(-x)  + (1-\sigma) \alpha'(x) $ where the first inequality follows as $ \alpha' $ is class~$ \calK_{e} $ and thus it is non-negative for non-negative arguments; the second inequality follows from the convexity of $ \alpha' $. By rearranging terms, we obtain $ -\alpha'(x)\leq \sigma (\alpha'(-x)-\alpha'(x)) $, and hence for $ \sigma = 1/2 $ and all $ x\geq 0 $ 
	\begin{align}
		\label{eq:time-varying CBF without input constraints 1 aux 2}
		-\alpha'(x)\leq\alpha'(-x).
	\end{align} 
	
	Now, we turn towards the actual proof of \eqref{eq:time-varying CBF without input constraints 1}. Let us define an extended version of $ \alpha_{2} $ as a convex extended class $ \calK_{e} $ function $ \alpha'_{2}: \bbR\rightarrow\bbR $ such that: (1) $ \alpha'_{2}(x) $ is an arbitrary continuous, convex and monotonously increasing continuation of $ \alpha_{2}(x) $ for all $ x<0 $; (2) $ \alpha'_{2}(x) = \alpha_{2}(x) $ for all $ x\in[0,A] $; and (3) $ \alpha'_{2}(x) = \alpha_{2}(A) + \alpha'_{1}(x\!-\!A) $ for all $ x\geq A $ where $ \alpha'_{1}(x) $ is some convex class~$ \calK $ function with $ \alpha'_{1}(x)\geq\alpha_{1}(x) $ for all $ x\geq0 $. 
	Next, we distinguish three cases, namely $ x_{1} \in [-A,0] $ with $ x_{1}+x_{2}\leq 0 $ (case~1a) and $ x_{1}+x_{2}\geq 0 $ (case~1b), and $ x_{1} \in [0,\infty) $ (case~2). Recall that $ x_{1}\in[-A,\infty) $ and $ x_{2}\in[0,A] $.
	
	\emph{Case~1a ($ x_{1} \in [-A,0] $ and $ x_{1}+x_{2}\leq0 $):} At first we note that since $ \alpha_{2} $ is convex it holds 
	\begin{align}
		\label{eq:time-varying CBF without input constraints 1 aux 3}
		\alpha_{2}(-(x_{1}\!+\! x_{2})) + \alpha_{2}(x_{2}) \stackrel{\eqref{eq:time-varying CBF without input constraints 1 aux 0}}{\leq} \alpha_{2}(-x_{1})
	\end{align}
	due to the superadditivity of $ \alpha_{2} $ and $ \alpha_{2}(0)=0 $. Next, we consider the left-hand side of~\eqref{eq:time-varying CBF without input constraints 1}. By employing that $ \alpha_{1}(-x)\leq -\alpha_{2}(x) $ for all $ x\in[0,A] $ and that $ \alpha'_{2} $ is convex, we obtain
	\begin{align*}
		\alpha_{1}(x_{1}) \!+\! \alpha_{2}(x_{2}) \!&\leq\! -\alpha_{2}(-x_{1}) \!+\! \alpha_{2}(x_{2}) \!\stackrel{\eqref{eq:time-varying CBF without input constraints 1 aux 3}}{\leq}\! -\alpha_{2}(-(x_{1}\!+\!x_{2})) \\
		&= -\alpha'_{2}(-(x_{1}+x_{2})) \stackrel{\eqref{eq:time-varying CBF without input constraints 1 aux 2}}{\leq} \alpha'_{2}(x_{1}+x_{2}).
	\end{align*}
	Here we employed that $ -(x_{1}+x_{2})\in[0,A] $ in case~1a.
	
	\emph{Case~1b ($ x_{1} \in [-A,0] $ and $ x_{1}+x_{2}>0 $):} Noting that $ x_{1} + A \geq 0 $, we derive by starting again with the left-hand side of~\eqref{eq:time-varying CBF without input constraints 1} that
	\begin{align*}
		\alpha_{1}(x_{1}) \!\!+\!\! \alpha_{2}(x_{2}) \!&\leq\! -\alpha_{2}(-x_{1}) \!\!+\!\! \alpha_{2}(x_{2}) \!=\!  -\alpha'_{2}(-x_{1}) \!\!+\!\! \alpha'_{2}(x_{2}) \\
		&\!\!\stackrel{\eqref{eq:time-varying CBF without input constraints 1 aux 1}}{\leq}\! -\alpha'_{2}(-x_{1}\!+\!x_{1}\!+\!A) \!+\! \alpha'_{2}(x_{1}\!+\!x_{2}\!+\!A) \\
		&= -\alpha'_{2}(A) \!+\! \alpha'_{2}(x_{1}\!+\!x_{2}\!+\!A).
	\end{align*}
	
	\emph{Case~2 ($ x_{1} \in [0,\infty) $):} Recall that by definition $ \alpha'_{2}(x) = \alpha_{2}(A) + \alpha'_{1}(x-A) $ for $ x\geq A $ with $ \alpha'_{1}(x)\geq\alpha_{1}(x) $ for all $ x\geq0 $. Thus, we have for $ x\geq 0 $ that
	\begin{align}
		\label{eq:time-varying CBF without input constraints 1 aux 4}
		\alpha_{1}(x) \leq \alpha'_{1}(x) = \alpha'_{2}(x+A) - \alpha_{2}(A).
	\end{align} 
	Furthermore, by employing that $ \alpha'_{2} $ is convex, we obtain
	\begin{align*}
		\alpha_{1}(x_{1}) \!+\! \alpha_{2}(x_{2}) \!&\stackrel{\eqref{eq:time-varying CBF without input constraints 1 aux 4}}{\leq}\! - \alpha_{2}(A) \!+\! \alpha'_{2}(x_{1}\!+\!A) \!+\! \alpha_{2}(x_{2}) \\ 
		&= \alpha'_{2}(x_{1}\!+\!A) \!-\! \alpha'_{2}(A) \!+\! \alpha'_{2}(x_{2}) \\ &\stackrel{\text{\eqref{eq:time-varying CBF without input constraints 1 aux 0}}}{\leq} -\alpha'_{2}(A) \!+\! \alpha'_{2}(A\!+\!x_{1}\!+\!x_{2}),
	\end{align*}
	which concludes the second case.
	
	Summarizing cases~1a, 1b and~2, we choose $ \beta $ in~\eqref{eq:time-varying CBF without input constraints 1} as 
	\begin{align*}
		\beta(x_{1}\!+\!x_{2}) = \begin{cases}
			\alpha'_{2}(x_{1}\!+\!x_{2}) &\text{if } x_{1}\!+\!x_{2}\leq0,  \\
			-\alpha'_{2}(A)\!+\!\alpha'_{2}(x_{1}\!+\!x_{2}\!+\!A) &\text{if } x_{1} \!+\! x_{2}> 0.
		\end{cases}
	\end{align*}
	As shown, $ \beta $ satisfies~\eqref{eq:time-varying CBF without input constraints 1}. Moreover, $ \beta $ is continuous, it holds $ \beta(0) = 0 $, and $ \beta $ is monotonously increasing since $ \alpha'_{2} $ is class $ \calK_{e} $; thus, also $ \beta $ is class~$ \calK_{e} $. This concludes the proof. 
\end{proof}

\begin{lemma}
	\label{lemma:time-varying CBF without input constraints 2}
	Let $ \alpha_{1}: \bbR \rightarrow \bbR $ be an extended class $ \calK_{e} $ function, and $ \alpha_{2}: \bbR_{\geq 0} \rightarrow \bbR $ a \emph{concave} class $ \calK $ function such that $ \alpha_{1}(-x)\leq -\alpha_{2}(x) $ for all $ x\in[0,A] $ and $ A>0 $. Then, there exists an extended class~$ \calK_{e} $ function $ \beta $ such that for all $ x_{1} \in [-A,\infty) $, $ x_{2} \in [0,A] $, it holds
	\begin{align}
		\label{eq:time-varying CBF without input constraints 2}
		\alpha_{1}(x_{1}) + \alpha_{2}(x_{2}) \leq \beta(x_{1}+x_{2}).
	\end{align}
	This even holds if $ A\rightarrow\infty $.
\end{lemma}
\begin{proof}
	The proof in the case of a concave function $ \alpha_{2} $ is more straightforward compared to the convex case. Before turning towards the actual proof, recall that the difference quotient $ D_{\alpha'}(x,y) := \frac{\alpha'(y)-\alpha'(x)}{y-x} $ of a concave function $ \alpha' $, where $ x<y $, is monotonously decreasing in both of its arguments. Thus, \eqref{eq:time-varying CBF without input constraints 1 aux 1} still holds, however only for non-positive $ c\leq 0 $. More precisely, it holds for all $ x<y $ and $ c\leq 0 $ that
	\begin{align}
		\label{eq:time-varying CBF without input constraints 2 aux 1}
		\alpha'(y)-\alpha'(x) \leq \alpha'(y+c)-\alpha'(x+c).
	\end{align}
	
	Now, we turn towards the actual proof of~\eqref{eq:time-varying CBF without input constraints 2}. To this end, we define again an extended version of $ \alpha_{2} $, however this time as a concave extended class $ \calK_{e} $ function $ \alpha'_{2}:\bbR\rightarrow\bbR $ such that $ \alpha'_{2}(x)=\alpha_{2}(x) $ for all $ x\geq 0 $. Next, we distinguish two cases, namely $ x_{1}\in[-A,0] $ (case~1) and $ x_{1}\in[0,\infty) $ (case~2). Recall that $ x_{2}\in[0,A] $.
	
	\emph{Case 1 ($ x_{1}\in[-A,0] $):} Consider the left-hand side of~\eqref{eq:time-varying CBF without input constraints 2}. By employing that $ \alpha_{1}(-x)\leq -\alpha_{2}(x) $ for all $ x\in[0,A] $ and that $ \alpha'_{2} $ is concave, we obtain
	\begin{align*}
		\alpha_{1}(x_{1}) \!&+\! \alpha_{2}(x_{2}) \!\leq\! -\alpha_{2}(-x_{1}) \!\!+\!\! \alpha_{2}(x_{2}) \!=\! -\alpha'_{2}(-x_{1}) \!\!+\!\! \alpha'_{2}(x_{2}) \\
		&\stackrel{\eqref{eq:time-varying CBF without input constraints 2 aux 1}}{\leq} -\alpha'_{2}(-x_{1} \!+\! x_{1}) \!+\! \alpha'_{2}(x_{1}\!+\!x_{2}) = \alpha'_{2}(x_{1}\!+\!x_{2}),
	\end{align*}
	where the last inequality is obtained by adding $ c=x_{1} $, which is non-positive by assumption, to the arguments of $ \alpha'_{2} $. For the case that $ x_{1} + x_{2}\geq 0 $, we note that the right-hand side is upper-bounded by
	\begin{align*}
		\alpha'_{2}(x_{1}+x_{2}) \leq \alpha_{1}(x_{1}+x_{2})+\alpha'_{2}(x_{1}+x_{2}).
	\end{align*}
	This observation is needed for the construction of the extended class~$ \calK_{e} $ function~$ \beta $.
	
	\emph{Case 2 ($ x_{1}\in[0,\infty) $):}  In this case, it always holds that $ x_{1}+x_{2}\geq 0 $. Thus, we directly obtain 
	\begin{align*}
		\alpha_{1}(x_{1}) \!+\! \alpha_{2}(x_{2}) &= \alpha_{1}(x_{1}) \!+\! \alpha'_{2}(x_{2}) \\
		&\leq \alpha_{1}(x_{1}\!+\!x_{2}) \!+\! \alpha'_{2}(x_{1}\!+\!x_{2}).
	\end{align*}
	
	Summarizing cases~1 and~2, we choose $ \beta $ in~\eqref{eq:time-varying CBF without input constraints 2} as
	\begin{align*}
		\beta(x_{1}\!+\!x_{2}) = 
		\begin{cases}
			\alpha'_{2}(x_{1}\!+\!x_{2}) &\text{if } x_{1}\!+\!x_{2}< 0, \\
			\alpha_{1}(x_{1}\!+\!x_{2}) \!+\! \alpha'_{2}(x_{1}\!+\!x_{2}) &\text{if } x_{1}\!+\!x_{2}\geq 0.
		\end{cases}
	\end{align*}
	As shown, $ \beta $ satisfies~\eqref{eq:time-varying CBF without input constraints 2}. Moreover, $ \beta $ is continuous, it holds $ \beta(0) = 0 $, and $ \beta $ is monotonously increasing as both $ \alpha_{1} $ and $ \alpha_{2}' $ are class $ \calK_{e} $; thus, also $ \beta $ is class~$ \calK_{e} $. This result even holds for $ A\rightarrow\infty $, as the construction of $ \beta $ does not rely on $ A $ being finite. This concludes the proof. 
\end{proof}

\begin{remark}
	To the best of our knowledge, the relations derived in Lemmas~\ref{lemma:time-varying CBF without input constraints 1} and~\ref{lemma:time-varying CBF without input constraints 2} have not been previously derived in the literature. Also in the compendium of comparison function results \cite{Kellett2014}, only results where $ \beta(x_{1}+x_{2}) $ constitutes a lower-bound can be found. 
\end{remark}

\begin{figure}[t]
	\centering
	\def\svgwidth{0.9\columnwidth}
	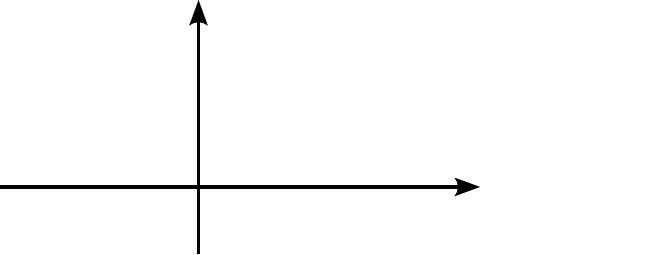
	\caption{The difference between the values of~$\alpha_{1}$ at arguments that differ with $x$ is not necessarily bounded for all $x_{1}$.}
	\label{fig:interpretation_alpha_condition}
\end{figure}

\begin{remark}
	\label{remark:necessity of alpha condition}
	Let us once more revisit the condition that $ \alpha_{1}(-x)\leq -\alpha_{2}(x) $ for all $ x\in[0,A] $, employed in both previous lemmas. This time, we take a more technical point of view. In particular, we note that this condition is even necessary in both lemmas. In order to see this, consider first Lemma~\ref{lemma:time-varying CBF without input constraints 1}. Let us assume that there exists an~$ x $ such that $ \alpha_{1}(-x)> -\alpha_{2}(x) $ for some $ x\in[0,A] $ and such that~\eqref{eq:time-varying CBF without input constraints 1} still holds. Specifically, let us consider $ x_{1} = -x $ and $ x_{2} = x $. Then, the left-hand side of~\eqref{eq:time-varying CBF without input constraints 1} is clearly positive, while the right-hand side is $ \beta(-x+x)=\beta(0) = 0 $ as $ \beta $ is class~$ \calK_{e} $. This, however, contradicts~\eqref{eq:time-varying CBF without input constraints 1}, and we conclude the necessity of the condition. For Lemma~\ref{lemma:time-varying CBF without input constraints 2}, necessity follows analogously. Intuitively, the condition ensures that the left-hand side of~\eqref{eq:time-varying CBF without input constraints 1} and~\eqref{eq:time-varying CBF without input constraints 2}, respectively, is negative whenever $ x_{1}+x_{2}<0 $, that is, whenever $ \beta $ is negative. Condition $ \alpha_{1}(-x)\leq -\alpha_{2}(x) $ is, however, not sufficient. To see this, apply the condition to the sum of $\alpha_{1}$ and $\alpha_{2}$, yielding 
	\begin{align*}
		\alpha_{1}(x_{1})+\alpha_{2}(x_{2})\leq\alpha_{1}(x_{1})-\alpha_{1}(x_{1}-x),
	\end{align*}
	where $x\coloneq x_{1}+x_{2}$. Here, the right-hand side is the difference between the values of~$\alpha_{1}$ at arguments that differ by $x$, see Figure~\ref{fig:interpretation_alpha_condition}. If the condition were sufficient, there would always exist an upper bound $\beta(x)$, as required by~\eqref{eq:class K function sum is upper-bounded}, that bounds this difference for any~$x_{1}$. Such $\beta$, however, only exists, when $\alpha$ is globally Lipschitz continuous. To avoid this limitation, we impose additional conditions on~$\alpha_{2}$, such as convexity and concavity.
\end{remark}
\begin{remark}
	Lemma~\ref{lemma:time-varying CBF without input constraints 2} provides a stronger result than Lemma~\ref{lemma:time-varying CBF without input constraints 1} where $ A $ is required to be finite. This is because the proofs of both lemmas build up on the monotonicity property of the difference quotient of $ \alpha_{2} $. In the case, when $ \alpha_{2} $ is convex, the monotonicity property only yields an upper-bound via~\eqref{eq:time-varying CBF without input constraints 1 aux 1} if $ c\geq 0 $, which invokes a finite constant $ A $ into the construction of $ \beta $. This is in contrast to the concave case, where the respective monotonicity property~\eqref{eq:time-varying CBF without input constraints 2 aux 1} requires only $ c\leq0 $, which circumvents the need for a finite $ A $. For details, refer to the proof of Lemma~\ref{lemma:time-varying CBF without input constraints 1}, case~1b, and Lemma~\ref{lemma:time-varying CBF without input constraints 2}, case~1.
\end{remark}

We are now prepared for the proof of Theorem~\ref{thm:time-varying CBF without input constraints}.

\begin{proof}
	As $ b(x) $ is locally Lipschitz continuous in $ x $ and $ \bm{\lambda}(t) $ satisfies Assumption~\ref{ass:lambda prop}, the continuity properties of $ B_{\bm{\lambda}(\cdot)} $ as required by Assumption~\ref{ass:continuity prop b} follow directly. 
	Thus, in order to show that $ B_{\bm{\lambda}(\cdot)} $ is a CBF, it remains to show that~\eqref{eq:def cbf dini time-varying} holds with respect to dynamics~\eqref{eq:dynamics}. To this end, recall that for the $ \Lambda $-shiftable CBF~$ b $, we have for all $ x\in\calC_{\Lambda} $ 
	\begin{align*}
		\sup_{u\in\calU} \left\{ db(x;f(x,u)) \right\} \geq -\alpha(b(x)).
	\end{align*}
	By adding $ -\alpha_{\lambda}(\bm{\lambda}(t)) $ to both sides, where $ \alpha_{\lambda} $ is as specified in the premises of the theorem, and applying Lemmas~\ref{lemma:time-varying CBF without input constraints 1} and~\ref{lemma:time-varying CBF without input constraints 2}, we obtain
	\begin{align}
		\label{eq:thm time-varying CBF without input constraints aux 0}
			&\sup_{u\in\calU} \left\{ db(x;f(x,u)) \right\} -\alpha_{\lambda}(\bm{\lambda}(t))  \nonumber\\
			&\quad\qquad\geq -\alpha(b(x)) - \alpha_{\lambda}(\bm{\lambda}(t)) \nonumber\\
			&\quad\qquad \hspace{-0.5cm}\stackrel{\text{Lem.~\ref{lemma:time-varying CBF without input constraints 1},\ref{lemma:time-varying CBF without input constraints 2}}}{\geq} -\beta(b(x)+\bm{\lambda}(t)).
	\end{align}
	Utilizing~\eqref{eq:lambda_dot condition}, we further derive
	\begin{align}
		\label{eq:thm time-varying CBF without input constraints aux 1}
		\sup_{u\in\calU} \left\{ db(x;f(x,u)) \right\} + d\bm{\lambda}(t;1) \geq -\beta(b(x)+\bm{\lambda}(t))
	\end{align}
	and we finally conclude that for all $ t\in\bbR_{\geq 0} $ and $ x\in\calC_{\Lambda} $
	\begin{align}
		\label{eq:thm time-varying CBF without input constraints aux 2}
		\begin{split}
			&\sup_{u\in\calU} \left\{ dB_{\bm{\lambda}(\cdot)}(t,x;1,f(x,u)) \right\} \\ 
			&\qquad= \sup_{u\in\calU} \left\{ db(x;f(x,u)) +  d\bm{\lambda}(t;1) \right\} \\
			&\qquad= \sup_{u\in\calU}\left\{ db(x;f(x,u)) \right\} + d\bm{\lambda}(t;1) \\ 
			&\qquad\stackrel{\eqref{eq:thm time-varying CBF without input constraints aux 1}}{\geq} -\beta(b(x)+\bm{\lambda}(t)).
		\end{split}
	\end{align}
	Thereby, we established condition~\eqref{eq:def cbf dini time-varying} for time-varying CBFs, which concludes the proof.
\end{proof}

A central step in the proof of Theorem~\ref{thm:time-varying CBF without input constraints} is ensuring the existence of an upper-bounding extended class~$\calK_{e}$ function~$\beta$ in~\eqref{eq:class K function sum is upper-bounded}. Also for the application of the theorem, the determination of~$\beta$ is essential. There are two approaches to this. First, an analytic construction of $\beta$ is presented in the proofs of Lemmas~\ref{lemma:time-varying CBF without input constraints 1} and~\ref{lemma:time-varying CBF without input constraints 2}. Second, $\beta$ can be determined numerically by solving the one-dimensional optimization problem
\begin{align*}
	\beta(x) \coloneq \sup_{\xi\in b(\calD)} \{\alpha(\xi) + \alpha_{\lambda}(x-\xi)\}.
\end{align*}
The computational complexity remains moderate since both the optimization variable and the argument of $\beta$ are one-dimensional, and the computation can be performed offline. In practice, it suffices to compute $\beta$ only in some neighborhood of zero, which corresponds to states, where $B_{\bm{\lambda}}$ is close to zero, and which are located close to the boundary of $\calC(t)$.

\subsection{Construction of Time-Varying Trajectories $ \bm{\lambda}(\cdot) $}
\label{subsec:lambda function construction guidelines}

The rate, at which a $ \Lambda $-shiftable CBF can be shifted through a trajectory $ \bm{\lambda} $, is determined by its extended class~$ \calK_{e} $ function $ \alpha $. This is a consequence of Theorem~\ref{thm:time-varying CBF without input constraints}, which requires
\begin{align}
	\label{eq:comlete lambda dot condition}
	d\bm{\lambda}(t;1) \geq -\alpha_{\lambda}(\bm{\lambda}(t)) \geq \alpha(-\bm{\lambda}(t)).
\end{align}
Provided with a $ \Lambda $-shiftable CBF, the condition suggests the following procedure to derive~$ \bm{\lambda} $:
\begin{enumerate}
	\item[(1)] Select a linear, convex or concave class~$ \calK $ function~$ \alpha_{\lambda} $ satisfying $ \alpha_{\lambda}(\xi)\leq -\alpha(-\xi) $ for all $ \xi\in[0,\Lambda] $. If $ \alpha $ is readily convex or concave on $ \bbR_{\leq 0} $, we directly choose $ \alpha_{\lambda}(\xi) \coloneq -\alpha(-\xi) $ (Corollary~\ref{corollary:time-varying CBF}). 
	\item[(2)] Choose any piecewise differentiable trajectory $ \bm{\lambda}: \bbR_{\geq 0} \rightarrow [0,\Lambda] $ with continuity properties specified in Assumption~\ref{ass:lambda prop} such that $ d\bm{\lambda}(t;1) \geq -\alpha_{\lambda}(\bm{\lambda}(t)) $ for all $ t\geq 0 $. Whenever $ \bm{\lambda} $ is differentiable, this reduces to $ \frac{\partial \bm{\lambda}}{\partial t} (t) \geq -\alpha_{\lambda}(\bm{\lambda}(t)) $. Intuitively, the smaller the value of $ \bm{\lambda} $, the smaller must be its decrease rate. 
	\item[(3)] The uniformly time-varying CBF is then, according to Theorem~\ref{thm:time-varying CBF without input constraints}, given by $ B_{\bm{\lambda}(\cdot)}(t,x) := b(x) + \bm{\lambda}(t) $.
\end{enumerate}

There are multiple approaches to address step~(2). If $ \bm{\lambda} $ is chosen as a piecewise linear function, its derivative is piecewise constant and the verification of~\eqref{eq:comlete lambda dot condition} becomes straightforward. Alternatively, if $ \bm{\lambda} $ shall be determined with maximal rate of change, the first order differential equation $ \dot{\bm{\lambda}}(t) = -\alpha_{\lambda}(\bm{\lambda}(t)) $ needs to be solved, which can be  done analytically or numerically depending on the choice of $ \alpha_{\lambda} $.

\subsection{Application to Time-Varying Constraints}
\label{sub_sec:application to time-varying constraints}

Uniformly time-varying CBFs are well-suited for ensuring the satisfaction of certain classes of time-varying constraints. Particularly constraints taking the form
\begin{align}
	\label{eq:uniformly time-varying constraint}
	x\in\calH(t)\coloneq \{ x\in\bbR^{n} \, | \, h(x) + \bm{\gamma}(t) \geq 0 \}
\end{align}
can be effectively handled, where $h$ defines the state-dependent component that is uniformly varied by $ \bm{\gamma}:\bbR_{\geq 0} \rightarrow [0,\Lambda] $. The constraint can be encoded into a uniformly time-varying CBF as follows:

\begin{enumerate}
	\item[(1)] First, determine a shiftable CBF $b$ satisfying $b(x)\leq h(x)$ for all $x\in\bbR^{n}$. The determination of a CBF $b$ based on a given constraint function $h$ is a well established problem in the literature. A review of available methods for determining $b$ as a shiftable CBF along with its associated extended class~$\calK_{e}$ function is provided in Section~\ref{sec:cbf and class K function construction}. 
	\item[(2)] Constraint~\eqref{eq:uniformly time-varying constraint} is encoded by the uniformly time-varying CBF $B(t,x)\coloneq b(x)+\bm{\lambda}(t)$ if there exists some $\bm{\lambda}$ satisfying~\eqref{eq:comlete lambda dot condition} such that $ \bm{\lambda}(t) \leq \bm{\gamma}(t) $ for all $ t\geq0 $. Such $\bm{\lambda}$, if it exists, can be constructed as outlined in Section~\ref{subsec:lambda function construction guidelines}. Overall, with these choices, it holds that if $B(t,x)\geq0$, then $x\in\calH(t)$; the forward control invariance of the zero super-level set of $B$ follows from Corollary~\ref{corollary:cbf invariance}.
\end{enumerate}

If $\bm{\gamma}$ satisfies the same continuity properties as~$\bm{\lambda}$ in Assumption~\ref{ass:lambda prop} and it is dynamically bounded in the sense that~\eqref{eq:comlete lambda dot condition} holds, then $\bm{\lambda}\equiv\bm{\gamma}$ is a valid choice in Step~(2). In this case, $\bm{\gamma}$ need not be known a priori and can instead be determined online. Otherwise, an estimate of $\bm{\gamma}$ may be required to adapt $\bm{\lambda}$ to the system's dynamic capabilities by ensuring~\eqref{eq:comlete lambda dot condition} in a predictive manner. 

Moreover, the proposed framework extends beyond uniformly time-varying constraints. In particular, constraints of the form $ \overline{\calH}(t) \coloneq \{ x \, | \, \bar{h}(t,x) \geq 0 \} $, where $ \bar{h}(t,x) \geq h(x) + \bm{\gamma}(t) $, can be handled using a uniformly time-varying CBF, provided the original constraint admits a (possibly conservative) uniformly time-varying approximation.

\subsection{Discussion}

The results developed in this section allow to decouple the time-variation of a CBF from the actual CBF synthesis. The construction of a uniformly time-varying CBF thereby consists of two parts, namely the construction of a CBF including the associated extended class~$ \calK_{e} $ function~$ \alpha $, and the design of a time-varying trajectory~$ \bm{\lambda} $, which is only confined by $ \alpha $ through~\eqref{eq:comlete lambda dot condition}. If the time-variation of the underlying constraint changes, only the time-varying function~$\bm{\lambda}$ needs to be redesigned, avoiding the recomputation of the entire CBF.  The conditions imposed on $ \bm{\lambda} $ allow for various realizations and can thereby account for different time-variations of the state-constraint without the need to change or recompute $ b $. As the computation of $ b $, despite the continued advances in the synthesis of CBFs, stays non-trivial, decoupling the synthesis of $ b $ from the time-variations is advantageous as it allows to reuse $ b $ in various contexts without the need for recomputation. In Section~\ref{sec:cbf and class K function construction}, we comment on the available methods for constructing shiftable CBFs~$ b $ and their associated extended class~$ \calK_{e} $ function~$ \alpha $. Note that decoupling the CBF construction may introduce conservatism, as the function~$b$, together with its associated extended class~$\calK_{e}$ function~$\alpha$, may permit faster variations according to~\eqref{eq:comlete lambda dot condition} than those prescribed by the chosen time-varying function~$\bm{\lambda}$. In return, the decoupled design substantially reduces computational effort. The degree of conservatism, if any, depends on the system dynamics as well as the non-holonomic and input constraints.

%% file: fig/illustrations/lambda_function_sketch.pdf_tex
\begingroup%
  \makeatletter%
  \providecommand\color[2][]{%
    \errmessage{(Inkscape) Color is used for the text in Inkscape, but the package 'color.sty' is not loaded}%
    \renewcommand\color[2][]{}%
  }%
  \providecommand\transparent[1]{%
    \errmessage{(Inkscape) Transparency is used (non-zero) for the text in Inkscape, but the package 'transparent.sty' is not loaded}%
    \renewcommand\transparent[1]{}%
  }%
  \providecommand\rotatebox[2]{#2}%
  \newcommand*\fsize{\dimexpr\f@size pt\relax}%
  \newcommand*\lineheight[1]{\fontsize{\fsize}{#1\fsize}\selectfont}%
  \ifx\svgwidth\undefined%
    \setlength{\unitlength}{405.5494716bp}%
    \ifx\svgscale\undefined%
      \relax%
    \else%
      \setlength{\unitlength}{\unitlength * \real{\svgscale}}%
    \fi%
  \else%
    \setlength{\unitlength}{\svgwidth}%
  \fi%
  \global\let\svgwidth\undefined%
  \global\let\svgscale\undefined%
  \makeatother%
  \begin{picture}(1,0.41776036)%
    \lineheight{1}%
    \setlength\tabcolsep{0pt}%
    \put(-0.00076755,0.2954051){\color[rgb]{0,0,0}\makebox(0,0)[lt]{\lineheight{1.25}\smash{\begin{tabular}[t]{l}$\Lambda$\end{tabular}}}}%
    \put(0,0){\includegraphics[width=\unitlength,page=1]{lambda_function_sketch.pdf}}%
    \put(0.01767614,0.03933085){\color[rgb]{0,0,0}\makebox(0,0)[lt]{\lineheight{1.25}\smash{\begin{tabular}[t]{l}$0$\end{tabular}}}}%
    \put(0.91548428,0.00135902){\color[rgb]{0,0,0}\makebox(0,0)[lt]{\lineheight{1.25}\smash{\begin{tabular}[t]{l}$t$\end{tabular}}}}%
  \end{picture}%
\endgroup%

%% file: fig/illustrations/interpretation_alpha_condition.pdf_tex
\begingroup%
  \makeatletter%
  \providecommand\color[2][]{%
    \errmessage{(Inkscape) Color is used for the text in Inkscape, but the package 'color.sty' is not loaded}%
    \renewcommand\color[2][]{}%
  }%
  \providecommand\transparent[1]{%
    \errmessage{(Inkscape) Transparency is used (non-zero) for the text in Inkscape, but the package 'transparent.sty' is not loaded}%
    \renewcommand\transparent[1]{}%
  }%
  \providecommand\rotatebox[2]{#2}%
  \newcommand*\fsize{\dimexpr\f@size pt\relax}%
  \newcommand*\lineheight[1]{\fontsize{\fsize}{#1\fsize}\selectfont}%
  \ifx\svgwidth\undefined%
    \setlength{\unitlength}{321.74690307bp}%
    \ifx\svgscale\undefined%
      \relax%
    \else%
      \setlength{\unitlength}{\unitlength * \real{\svgscale}}%
    \fi%
  \else%
    \setlength{\unitlength}{\svgwidth}%
  \fi%
  \global\let\svgwidth\undefined%
  \global\let\svgscale\undefined%
  \makeatother%
  \begin{picture}(1,0.37958849)%
    \lineheight{1}%
    \setlength\tabcolsep{0pt}%
    \put(0,0){\includegraphics[width=\unitlength,page=1]{interpretation_alpha_condition.pdf}}%
    \put(0.21219662,0.33514475){\color[rgb]{0,0,0}\makebox(0,0)[lt]{\lineheight{1.25}\smash{\begin{tabular}[t]{l}$\alpha_1$\end{tabular}}}}%
    \put(0.48380517,0.0550689){\color[rgb]{0,0,0}\makebox(0,0)[lt]{\lineheight{1.25}\smash{\begin{tabular}[t]{l}$x_1$\end{tabular}}}}%
    \put(0.33472652,0.0550658){\color[rgb]{0,0,0}\makebox(0,0)[lt]{\lineheight{1.25}\smash{\begin{tabular}[t]{l}$x_1\!-\!x$\end{tabular}}}}%
    \put(0,0){\includegraphics[width=\unitlength,page=2]{interpretation_alpha_condition.pdf}}%
    \put(0.44263352,0.11574127){\color[rgb]{0,0,0}\makebox(0,0)[lt]{\lineheight{1.25}\smash{\begin{tabular}[t]{l}$x$\end{tabular}}}}%
    \put(0.51171411,0.16025767){\color[rgb]{0,0,0}\makebox(0,0)[lt]{\lineheight{1.25}\smash{\begin{tabular}[t]{l}$\alpha_{1}(x_1)\!-\!\alpha_{1}(x_1\!-\!x)$\end{tabular}}}}%
    \put(0,0){\includegraphics[width=\unitlength,page=3]{interpretation_alpha_condition.pdf}}%
  \end{picture}%
\endgroup%

%% file: 03-relation-of-lambda-shiftable-CBFs-and-CLFs.tex
\section{Relation of CLFs and Shiftable CBFs in the Dini Sense}
\label{sec:relation of lambda shiftable cbf and clf}

Establishing the relation between (shiftable) CBFs and Control Lyapunov Functions (CLF) allows to take advantage of the synthesis methods developed for CLFs in the construction of uniformly time-varying CBFs. Such methods have a long history in the literature, and for some systems even analytic CLFs could be derived. 

CBFs possess properties similar to those of Control Lyapunov Functions. Conceptually, a CBF $ b $ guarantees that for any state $ x\in\calC $, there exists a control input $ \bm{u} $ such that the rate of decrease of $ b $ along trajectory $ \bm{\varphi}(\cdot;x_{0},\bm{u}) $ is lower-bounded by a prescribed rate $ \alpha $, see~\eqref{eq:def cbf 1}. For $ x\in\calD\smallsetminus\calC $, the condition is stricter requiring a control input that ensures a minimum rate of ascend on $ b $, which leads to the asymptotic stabilization of $ \calC $ \cite[Prop.~2]{Ames2017}. A property similar to the latter one also holds for CLFs in the Dini sense.

\begin{definition}[CLF in the Dini sense; \cite{Clarke2011}]
	\label{def:clf_dini}
	Let $ V: \bbR^{n}\rightarrow\bbR_{\geq 0} $ be Lipschitz-continuous, proper in the sense that all level sets $ \{x \, | \, V(x)\leq \lambda\} $ are compact for every $ \lambda $, and positive definite. Then $ V $ is called a \emph{Control Lyapunov Function in the Dini sense} on $ \calD $ with respect to \eqref{eq:dynamics}, where $ \calD\subseteq\bbR^{n} $ is some neighborhood of the origin, if there exists a class $ \calK $ function $ \gamma $ such that for all $ x\in\calD $
	\begin{align}
		\label{eq:def clf 1}
		\sup_{u\in\calU}\left\{ dV(x;f(x,u)) \right\} \leq -\gamma(V(x)).
	\end{align}
\end{definition}

\begin{theorem}\textbf{\emph{(Relation of $ \Lambda $-shiftable CBFs and CLFs)}}
	\label{thm:clf and cbf}
	Let $ V $ be a CLF  in the Dini sense on $ \calD\subseteq\bbR^{n} $ with respect to~\eqref{eq:dynamics}, and let $ \Lambda_{\text{max}} $ be the maximum value on its largest closed level set contained in domain $ \calD $, that is $ \Lambda_{\text{max}} := \max\{\lambda \,|\, \calC^{V}_{\lambda}:=\{x \, | \, V(x) \leq \lambda \} \subseteq \calD\} $. Furthermore, consider 
	\begin{align}
		\label{eq:thm:clf based cbf}
		b(x) := -V(x) + b_{c},
	\end{align}
	where $ b_{c}\in[0,\Lambda_{\text{max}}) $ is some constant. Then $ b $ is a $ \Lambda $-shiftable CBF for any $ \Lambda\in(0,\Lambda_{\text{max}}-b_{c}] $ if $ \Lambda_{\text{max}} $ is finite, and otherwise for any $ \Lambda\in (0,\infty) $. Moreover, if $ \calD=\bbR^{n} $, then $ b $ is a $ \Lambda $-shiftable CBF with $ \Lambda\in(0,\infty) $. 
\end{theorem}

\begin{proof}
	By Definition~\ref{def:lambda shiftable CBF}, $ b $ is a $ \Lambda $-shiftable CBF if for all $ x\in\calC_{\Lambda} $
	\begin{align}
		\label{eq:thm:clf based cbf aux 0}
		\begin{split}
			\sup_{u\in\calU}\left\{ db(x;f(x,u)) \right\} &\stackrel{\eqref{eq:thm:clf based cbf}}{=} \sup_{u\in\calU}\left\{-dV(x;f(x,u)) \right\} \\
			&\,\geq -\alpha(-V(x)+b_{c})
		\end{split}
	\end{align} 
	for some extended class $ \calK_{e} $ function $ \alpha $. 
	Starting with the left-hand side of the latter equation, we derive
	\begin{align}
		\label{eq:thm:clf based cbf aux 1}
		\begin{split}
			\sup_{u\in\calU}\left\{-dV(x;f(x,u)) \right\} &\stackrel{\eqref{eq:def clf 1}}{\geq} \gamma(V(x)) \geq \alpha(V(x)) \\ 
			&\geq \alpha(V(x)-b_{c}),
		\end{split} 
	\end{align} 
	where $ \alpha $ is a suitable extended class $ \calK_{e} $ function chosen as $ \alpha(x) = \left\{ \begin{smallmatrix}
		\gamma(x) & \text{if } x\geq 0, \hfill \\
		-\gamma(-x) & \text{if } x<0 \hfill
	\end{smallmatrix}\right. $. Next, we note that for this choice of $ \alpha $, we have $ \alpha(x) = -\alpha(-x) $. Thus, we further obtain for the right-hand side of \eqref{eq:thm:clf based cbf aux 1} that $ \alpha(V(x)-b_{c}) = -\alpha(-V(x)+b_{c}) $. Substituting this into~\eqref{eq:thm:clf based cbf aux 1} yields~\eqref{eq:thm:clf based cbf aux 0}. As~\eqref{eq:def clf 1} only holds on $ \calD $, the largest closed super-level set of $ b $, where the CBF gradient condition~\eqref{eq:thm:clf based cbf aux 0} holds, is then given by $ \calC_{\Lambda_{\text{max}}-b_{c}} = \{x\, | \, b(x) \geq -\Lambda_{\text{max}}+b_{c}\} $ if $ \Lambda_{\text{max}} $ is finite, and otherwise by $ \lim_{\Lambda\rightarrow\infty} \calC_{\Lambda} = \bbR^{n} $. Thus, we conclude that $ b $ is $ \Lambda $-shiftable for any $ \Lambda\in(0,\Lambda_{\text{max}}-b_{c}] $ if $ \Lambda_{\text{max}} $ is finite, and otherwise for any $ \Lambda \in (0,\infty) $.
\end{proof}

%% file: 05-some-guidelines-on-the-construction-of-alpha-functions.tex
\section{On the Construction of Shiftable CBFs and the Associated Extended Class $ \calK_{e} $ Function $ \alpha $}
\label{sec:cbf and class K function construction}

Commonly, CBF synthesis schemes yield an associated extended class~$ \calK_{e} $ function, denoted by $ \alpha $, along with the CBF. Yet, some of the synthesis schemes have limitations that are important to take into account. 

As such, \emph{analytic construction methods} as the design of higher-order CBFs via backstepping~\cite{Tan2021,Xiao2021b} do not directly yield $ \alpha $ in the presence of input constraints. Similar considerations hold for analytically constructed CLFs, or CLFs found in the literature, where $ \gamma $ is generally not known. The design of $ \alpha $ is then conducted via the construction of an auxiliary controller. In particular, for a given CBF $ b $, a controller $ k: \calD\rightarrow\calU $ needs to be constructed such that either $ -db(x;f(x,k(x))) $ is class~$ \calK_{e} $, or can be at least upper-bounded by an extended class~$ \calK_{e} $ function, which constitutes $ \alpha $. The approach is analogous for CLFs, which we illustrate for the derivation of $ \gamma $ via an LQR controller.

\begin{example}
	\label{example:lqr}
	Consider the linear system $ \dot{x}=Ax+Bu $ and cost function $ J(x_{0},u) = \min_{u} \int_{0}^{\infty} ||x(\tau)||_{Q}^{2} + ||u(\tau)||_{R}^{2} \, d\tau $, where $ x(\tau)\coloneq\bm{\varphi}(\tau;x_{0},u) $, $ Q=C^{T}C $ positive semi-definite and $ R $ positive definite. The cost functional is minimized for $ u^{\ast}(t) = -R^{-1} B^{T}  P \,x(\tau) $ and it holds $ J^{\ast}(x_{0})=x_{0}^{T} P x_{0} $, where $ P $ is the positive definite solution to the algebraic Riccati equation $ PA + A^{T}P - PBR^{-1}B^{T}P + Q = 0 $; the pair $ (A,B) $ is assumed to be controllable and $ (A,C) $ observable~\cite{Kalman1964}. It is well known that $ V(x)\coloneq J^{\ast}(x) $ is a CLF, and  based on $ u^{\ast} $ we derive the associated class~$ \calK $ function $ \gamma $ for $ V $ as
	\begin{align*}
		\dot{V}(x) &= 2x^{T} P \dot{x} = x^{T}A^{T}Px+x^{T}PAx + 2x^{T}PBu^{\ast}\\
		&= x^{T}( A^{T}P + PA - 2PBR^{-1}B^{T}P) x \\  
		&= -x^{T}(Q+PBR^{-1}B^{T}P)x = -x^{T}\widetilde{Q}x  \\
		&\leq -\lambda_{\text{min}}(\widetilde{Q})||x||^{2} \leq -\frac{\lambda_{\text{min}}(\widetilde{Q})}{\lambda_{\text{max}}(P)} V(x) \eqcolon -\gamma(V(x)),
	\end{align*}
	where $ \widetilde{Q}\coloneq Q+PBR^{-1}B^{T}P $ is positive definite, and $ \lambda_{\text{min}} $ and $ \lambda_{\text{max}} $ denote the smallest and largest eigenvalues, respectively. Matrix~$ R $ can be adjusted to account for input constraints.
\end{example}

Alternatively, also Sontag's formula~\cite{Sontag1989} can be used as an explicit feedback control law in the construction of~$ \gamma $, and the analogue for CBFs~\cite{Wieland2007} for the construction of~$ \alpha $. 

Furthermore, not all CBF synthesis methods are suitable for the construction of shiftable CBFs. Some are confined to CBFs with domain $ \calD = \calC $, which relates them closer to barrier certificates~\cite{Prajna2004} than to Lyapunov-like functions. Methods suitable for the numerical synthesis of shiftable CBFs include methods based on reachability~\cite{Wiltz2025b}, SOS~\cite{Dai2023} and learning techniques~\cite{Lindemann2024a,Chen2024c}. All of the methods are optimization-based and adjust the CBF to its associated extended class~$ \calK_{e} $ function, thereby resulting in a less restrictive lower-bound in~\eqref{eq:comlete lambda dot condition}. 
In~\cite{Wiltz2025b,Lindemann2024a}, the extended class~$ \calK_{e} $ function can be freely chosen as a design parameter, while~\cite{Dai2023,Chen2024c} consider it to be linear. For systems exhibiting equivariances, the computational effort for numeric methods can be reduced by exploiting these in the synthesis \cite{Wiltz2025c}. Beyond numeric synthesis methods, an analytic method for Euler-Lagrange systems is~\cite{Cortez2022a}, and for feedback linearizable systems without input constraints~\cite{Cohen2024b}. Since CLFs directly give rise to shiftable CBFs, as previously noted, we can also draw upon the results from the corresponding literature. CLFs in the Dini sense are derived via an MILP in~\cite{Baier2018}. Other approaches include SOS~\cite{Papachristodoulou2002,Tan2004}, Hamilton-Jacobi Bellman equations~\cite{Homer2019} or are based on learning techniques~\cite{Chang2019,Ravanbakhsh2019,Abate2021a}. Methods that are designed for the synthesis of Lyapunov functions are applicable as well, but require the system to be prestabilized.

%% file: 10-simulation.tex
\section{Numerical Examples}
\label{sec:simulation}

We demonstrate the application of our results through several illustrative examples. Specifically, we consider:
\begin{enumerate}
	\item a three-wheeled omni-directional robot as example for a first-order system tracking a sequence of waypoints within prescribed time implemented through time-varying sets;
	\item a disturbed mechanical pendulum as example for a second order system subject to time-varying constraints on its angle of excitation;
	\item a 3d-double integrator subject to gravity (``linearized quadcopter'') as example for a linear system;
	\item a unicycle as example for a nonholonomic system tracking a sequence of waypoints within prescribed time implemented through time-varying sets;
	\item avoidance of a time-varying obstacle by various systems including single and double integrators, kinematic bicycle models and a unicycle, each of them subject to input constraints. 
\end{enumerate}
Notably, the last two examples employ nonsmooth uniformly time-varying CBFs in the Dini sense, whereas the first three use smooth ones.

\begin{figure*}[th]
	\centering

	\begin{adjustbox}{width=\textwidth}
	\addtocounter{figure}{-1}
	
	\parbox[t]{0.33\linewidth}{
		\centering
		\begin{subfigure}[b]{0.9\linewidth}
			\centering
			\includegraphics[width=0.9\linewidth]{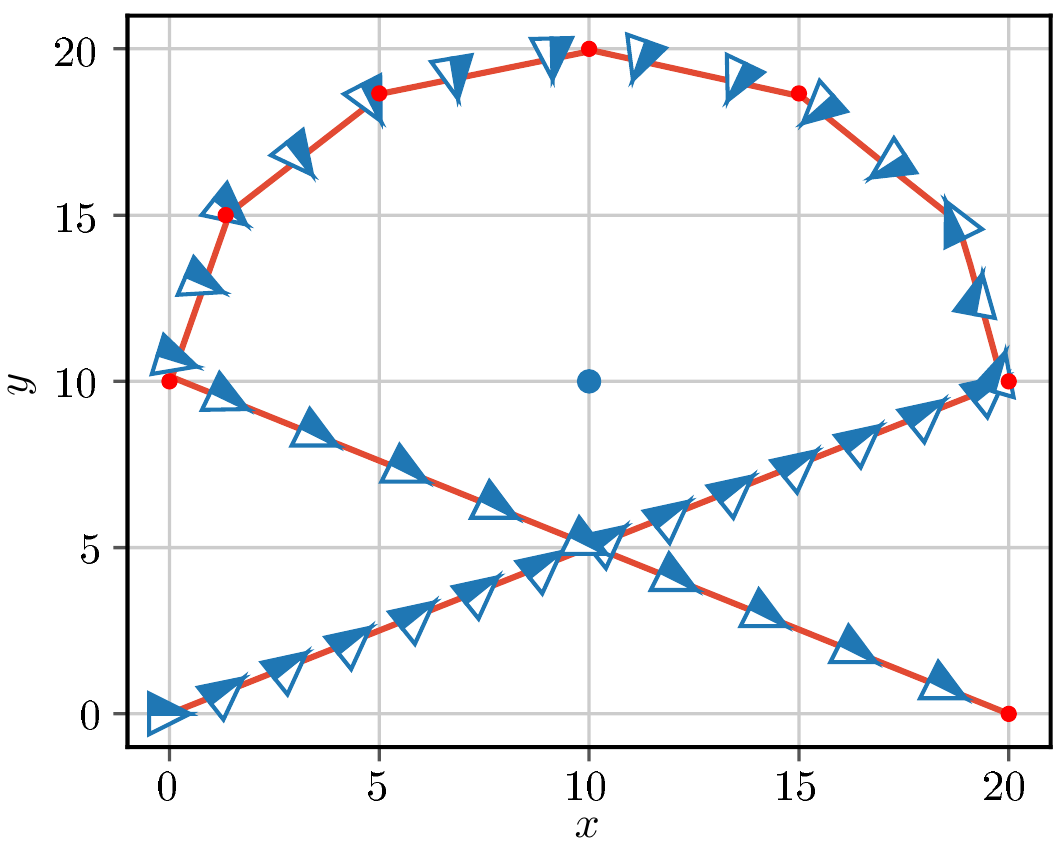}
			\caption{State trajectory including robot orientation (arrows), waypoints (red dots) and the center point as point of interest (blue dot).}
			\label{sub_fig:sim_omni_traj}
			\vspace{0.2cm}
		\end{subfigure}
		\linebreak
		\begin{subfigure}[b]{0.9\linewidth}
			\centering
			\includegraphics[width=1.05\linewidth]{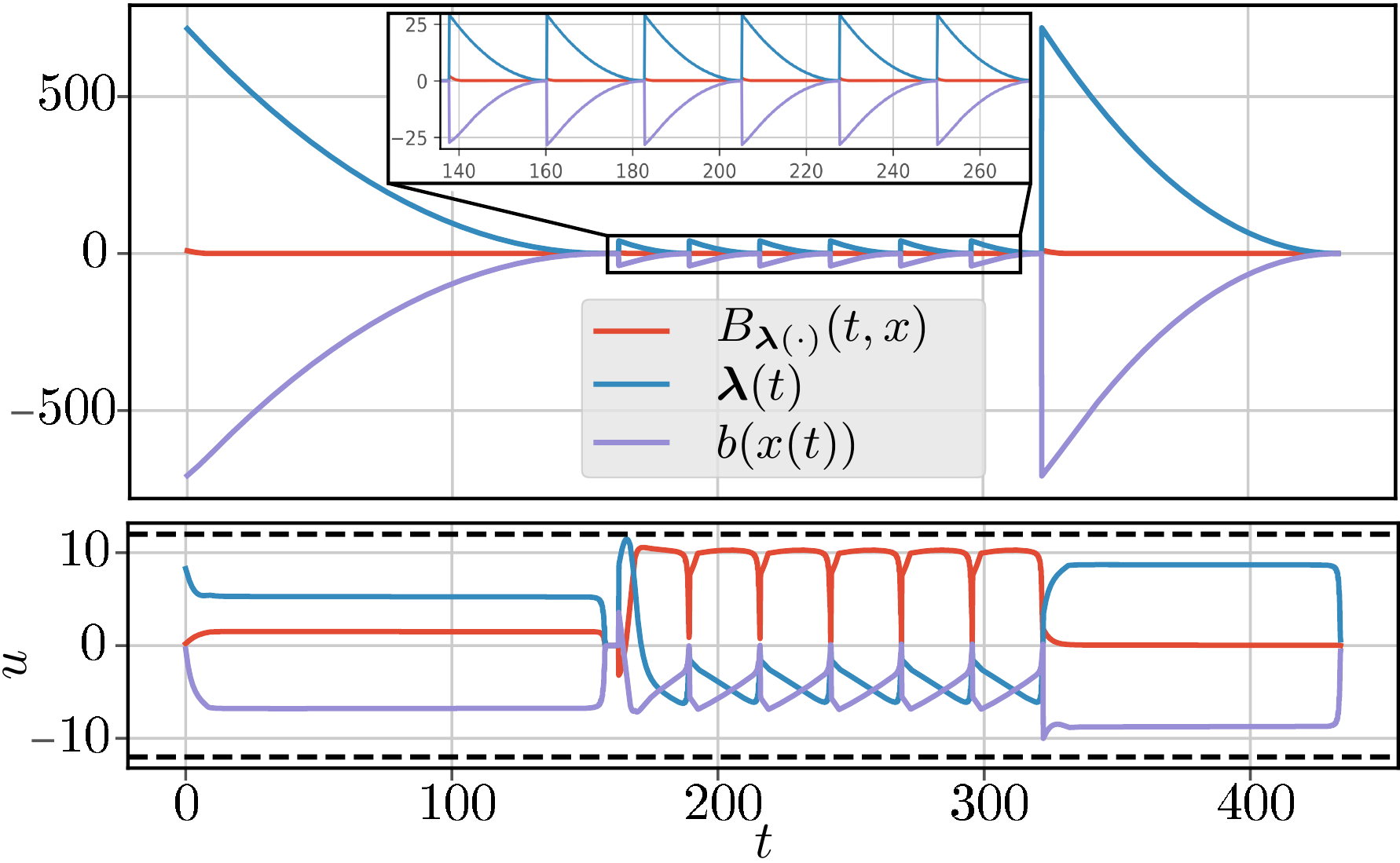}
			\caption{Value functions and inputs over time; the input constraints are indicated by the dashed lines.}
			\label{sub_fig:sim_omni_cbf_traj_and_inputs}
		\end{subfigure}
		\captionof{figure}{Example three-wheeled omni-directional robot.}
		\label{fig:sim_omni}
	}

	\hfill
	
	\parbox[t]{0.33\linewidth}{
		\centering
		\begin{subfigure}[b]{0.9\linewidth}
			\centering
			\includegraphics[width=0.9\linewidth]{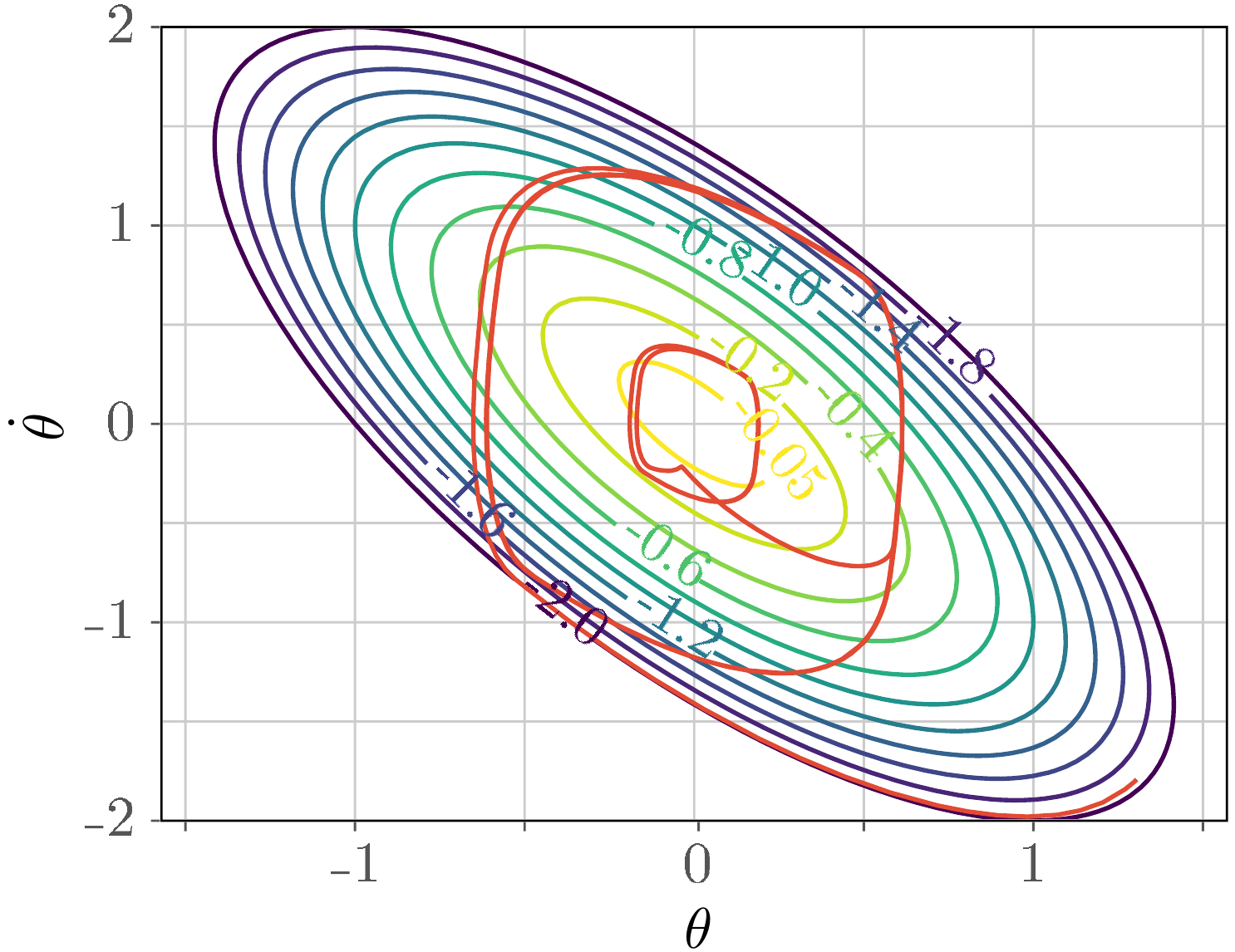}
			\caption{Level sets of $ b $ and phase plot for $ t\in[0,20] $.}
			\label{sub_fig:sim_pendulum_phase_plot}
			\vspace{2\baselineskip}
		\end{subfigure}
		\linebreak
		\begin{subfigure}[b]{0.9\linewidth}
			\vspace{-\baselineskip}
			\centering
			\includegraphics[width=1.05\linewidth]{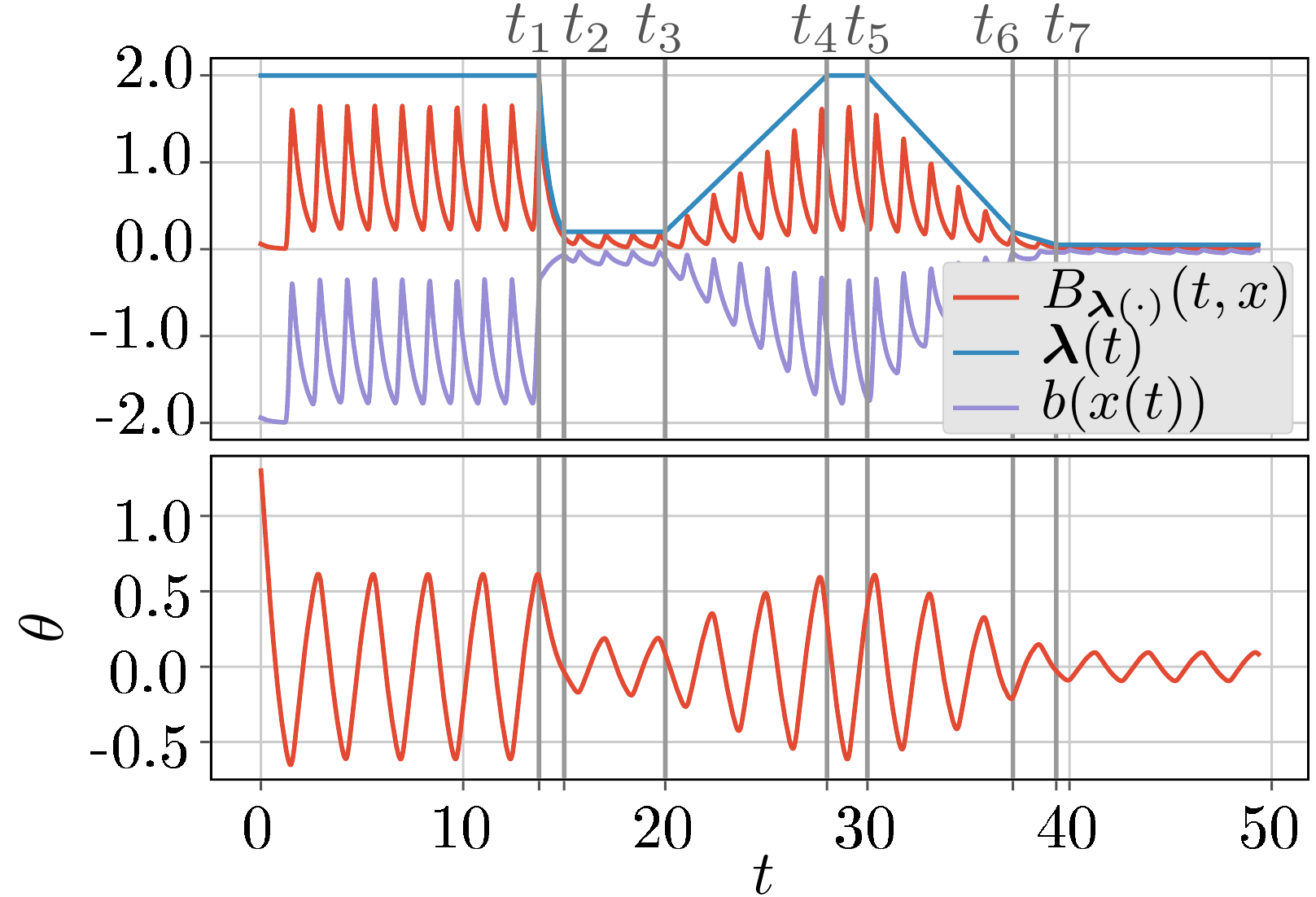}
			\caption{Value functions and angle of excitation over time. \vspace{\baselineskip}}
			\label{sub_fig:sim_pendulum_theta}
		\end{subfigure}

		\captionof{figure}{Example disturbed pendulum.}
		\label{fig:sim_pendulum}
	}

	\hfill
	
	\parbox[t]{0.33\linewidth}{
		\centering
		\begin{subfigure}[b]{0.9\linewidth}
			\centering
			\includegraphics[width=0.95\linewidth]{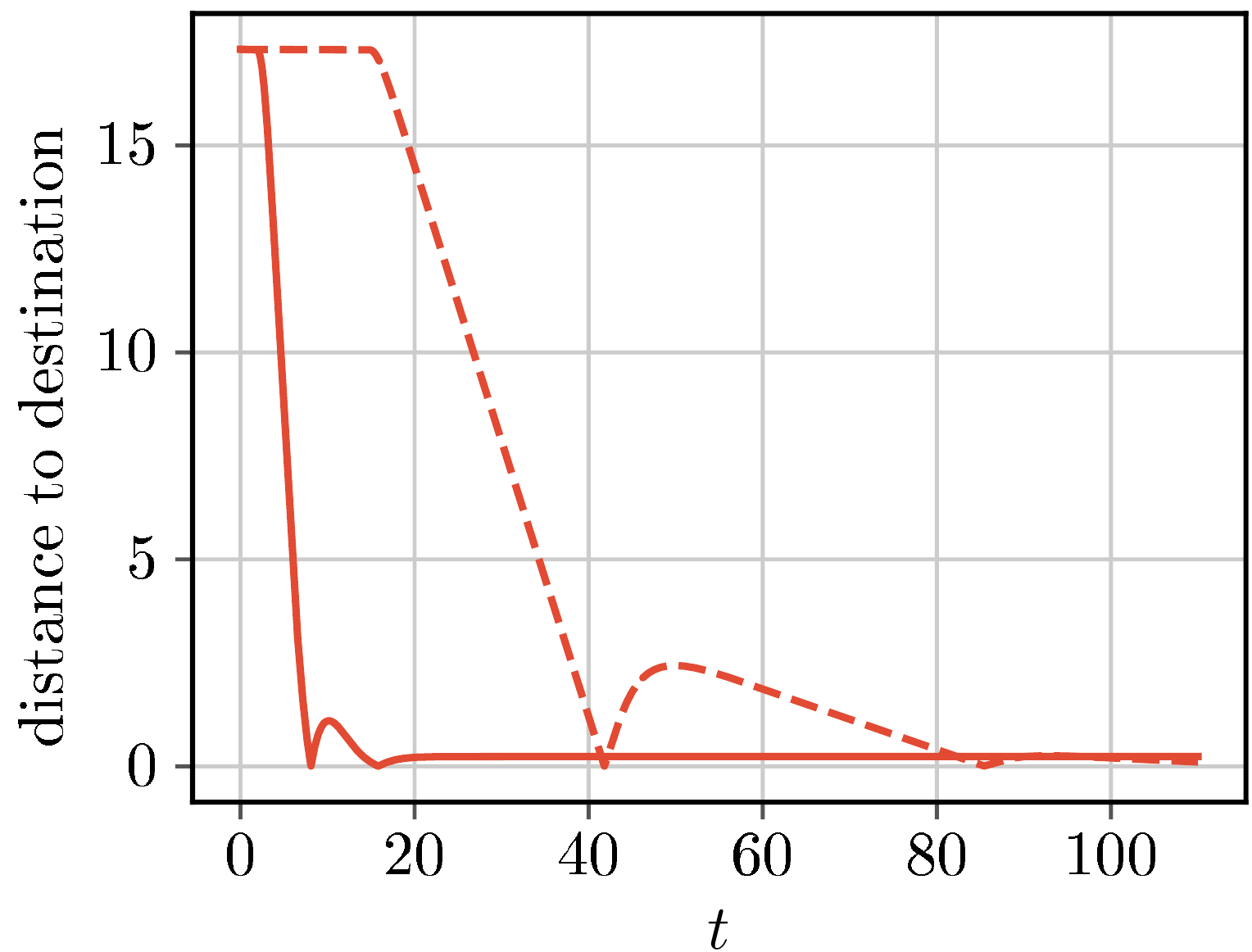}
			\vspace{-0.2cm}
			\caption{\vspace{3.5\baselineskip}}
			\label{sub_fig:sim_quadcop_dist}
		\end{subfigure}
		\linebreak
		\begin{subfigure}[b]{0.9\linewidth}
			\vspace{-1.3\baselineskip}
			\centering
			\includegraphics[width=1.0\linewidth]{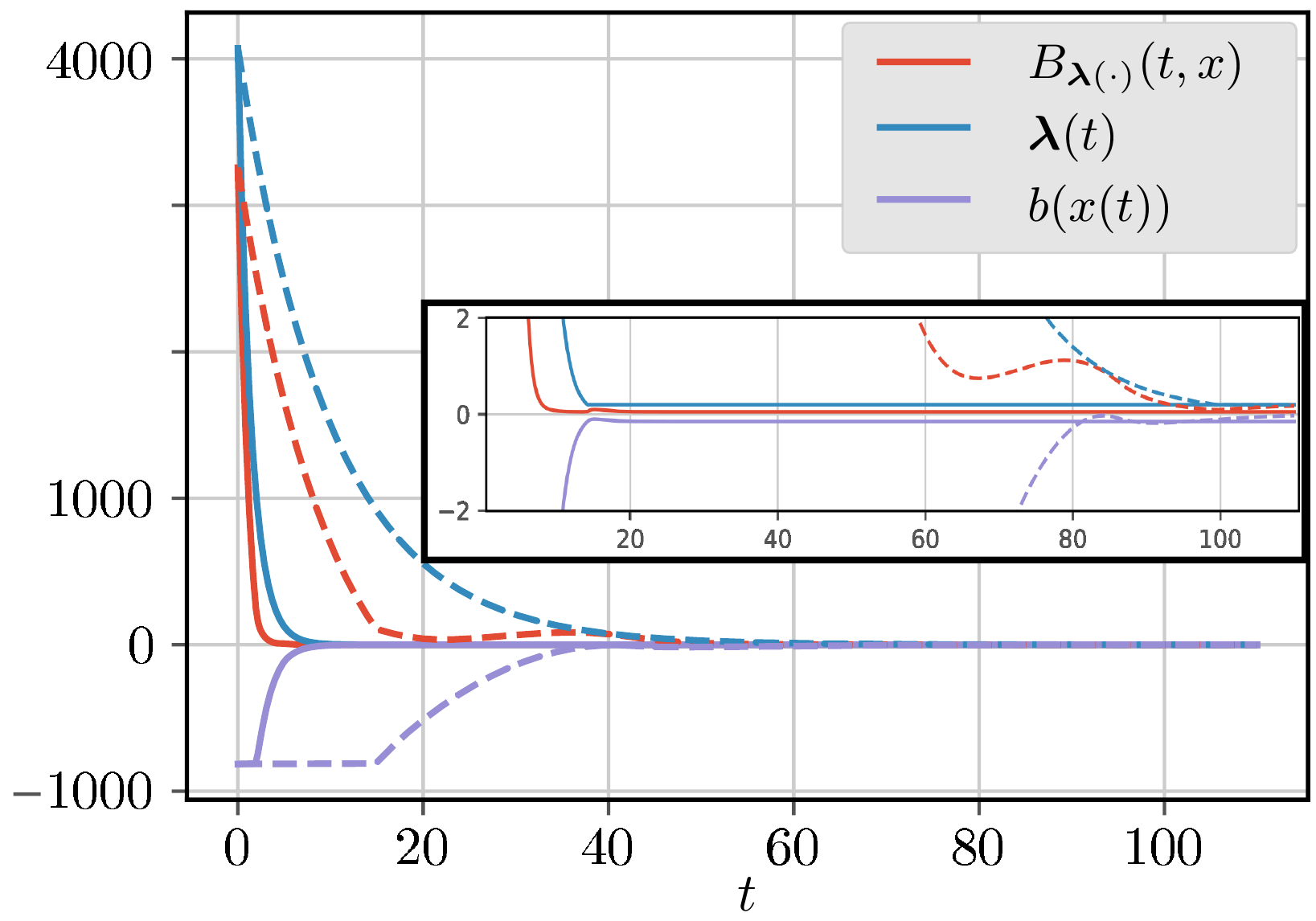}
			\vspace{-0.55cm}
			\caption{}
			\label{sub_fig:sim_quadcop_cbf_traj}
		\end{subfigure}
		\captionof{figure}{Linear system: comparison of an analytically computed extended class~$ \calK_{e} $ function $ \alpha $ (dashed lines) and the numerically computed least conservative one (solid).}
		\label{fig:sim_quadcop}
	}

	\end{adjustbox}

\end{figure*}

\subsection{First Order System: Omni-Directional Mobile Robot}
\label{subsec:sim_omni}

Consider the kinematic model of a three-wheeled omni-directional robot~\cite{Liu2008}
\begin{align*}
	\dot{x} = 
	\left[ \begin{smallmatrix}
		\cos(\rho) & -\sin(\rho) & 0 \\
		\sin(\rho) & \cos(\rho) & 0 \\
		0 & 0 & 1
	\end{smallmatrix} \right]
	\left(B^{T}\right)^{-1} R u = \Gamma(x)\, \left(B^{T}\right)^{-1}  R \,u,
\end{align*}
where $ B= \left[ \begin{smallmatrix}
	0 & \cos(\pi/6) & -\cos(\pi/6) \\ 
	-1 & \sin(\pi/6) & \sin(\pi/6) \\
	L & L & L
\end{smallmatrix} \right] $
with $ L = 0.2 $ as the radius of the robot body, and $ R=0.02 $ as wheel radius; the state vector $ x = [p_x, p_y, \rho]^{T} $ consists of position $ p = [p_x, p_y]^{T} $ and orientation $ \rho $, and input $ u\in\bbR^{3} $ is the angular velocity of the three wheels, which serve as control input. The system is subject to input constraints $ ||u||_{\infty}\leq u_\text{max} $. The task of the robot is to track a sequence of waypoints $ \{q_i\} $, $q_{i}\in\bbR^{2}$, within prescribed time~$ T_{i} $ while pointing towards a point of interest. Waypoint tracking is achieved using uniformly time-varying CBFs that encode the constraints
\begin{align}
	\label{eq:time-varying constraint waypoint tracking}
	x\!\in\!\calH_{i}(t) \!\coloneq\! \big\{ x \!\in\! \bbR^{3} \, \big| \, -\!||p-q_{i}||_{Q}^{2}\!+\!\bm{\lambda}_{i}(t)^{2} \!\geq\! 0 \big\}
\end{align}
with $t\!\leq\! T_{i}$, where $ Q\in\bbR^{2\times2} $ is some positive-definite matrix, and $\bm{\lambda}_{i}\!:\![0,T_{i}]\!\rightarrow\!\bbR$ with $\bm{\lambda}_{i}(T_{i})\!=\!0$ is chosen to be as large as possible at each time while ensuring that the constraint can still be satisfied under the given input constraints. The robot's orientation is controlled by a baseline controller.

As candidate for a shiftable CBF, we choose, in accordance with the time-invariant part in~\eqref{eq:time-varying constraint waypoint tracking},
\begin{align}
	\label{eq:omni_b}
	b(x) = -||p||_{Q}^{2} + r^{2},
\end{align} 
where $ r\geq 0 $; we account for the shift by~$q_{i}$ later on. This is a valid CBF candidate since its time-derivative depends explicitly on the control input for all $ x\in\bbR^{3} $. It remains to show that there exists an extended class~$ \calK_{e} $ function $ \alpha $ such that
\begin{align}
	\label{eq:omni_b_dot}
	\dot{b}(x) = -2x^{T} 
	\left[\begin{smallmatrix}
		Q & \bm{0} \\ \bm{0} & \bm{0}
	\end{smallmatrix}\right]
	\Gamma(x)\, \left(B^{T}\right)^{-1} R\, u \geq -\alpha(b(x))
\end{align}
for some control input $ u $ satisfying the input constraint. To this end, assuming that such $ \alpha $ exists, choose the state-feedback controller 
\begin{align*}
	u(x) = \tfrac{\alpha(b(x))}{2R (r^{2}-b(x))} B^{T} \left[\begin{smallmatrix}
		\bm{I} & 0 \\
		0 & 0
	\end{smallmatrix}\right] \Gamma^{-1}\!(x) \left[\begin{smallmatrix}
		p \\ 0
	\end{smallmatrix}\right]
\end{align*}
which satisfies~\eqref{eq:omni_b_dot}. Let now $ r $ in~\eqref{eq:omni_b} be zero. Then, for incorporating the input constraints, we derive
\begin{align*}
	&||u(x)||_{\infty} \leq \tfrac{|\alpha(b(x))|}{2R \, |b(x)|} \left|\left| B^{T} \left[\begin{smallmatrix}
		\bm{I}_{2} & \bm{0} \\
		\bm{0} & \bm{0}
	\end{smallmatrix}\right] \Gamma^{-1}\!(x) \right|\right|_{\infty} \, ||p||_{\infty} \\
	&\qquad\leq \tfrac{|\alpha(b(x))|}{2R \, |b(x)|} \sqrt{2}\, ||p||_\infty \leq \tfrac{|\alpha(b(x))|}{\sqrt{2}R |b(x)|} \sqrt{\tfrac{|b(x)|}{\lambda_{\text{min}}(Q)}} \stackrel{!}{\leq} u_{\text{max}},
\end{align*}
where we exploited in the second last inequality that $ p^{T}Qp \geq \lambda_{\text{min}}(Q) ||p||_{2}^{2} \geq \lambda_{\text{min}}(Q) ||p||_{\infty}^{2} $ with $ \lambda_{\text{min}}(Q) $ denoting the smallest eigenvalue of $ Q $; the derived terms need to be upper-bounded by $ u_{\text{max}} $ in order to account for the input constraints. An extended class~$ \calK_{e}$ function satisfying the last inequality is
\begin{align*}
	\alpha(b(x)) = \sqrt{2} \; \text{sgn}(b(x))\, R \, u_{\text{max}}\, \sqrt{\lambda_{\text{min}}(Q) \, |b(x)|}.
\end{align*}
Thereby, \eqref{eq:omni_b_dot} is satisfied for all $ x\in\bbR^{3} $ as well, and we conclude that $ b $ is $ \Lambda $-shiftable with $ \Lambda \rightarrow\infty $. We point out that the associated $ \alpha $ here depends on a square root and is thereby nonlinear, which is in contrast to the common choice of $ \alpha $ as a linear function in the literature. Specifically, the choice of $ \alpha $ as a linear function would only account for the input constraint for finite $ \Lambda $. At last, we note that $ b $ can be centered around any waypoint $ q_{i} $ by replacing $ p $ with $ \Delta p \coloneq p - q_{i} $.
From here on, Theorem~\ref{thm:time-varying CBF without input constraints} can be directly applied to the construction of uniformly time-varying CBFs. 

In this example, we assume $ u_{\text{max}} = 12 $ and choose $ Q=\bm{I} $. Next, we choose $ \alpha_{\lambda}(\xi) \coloneq -\alpha(-\xi) $ for $ \xi\in\bbR_{\geq0} $ as concave class~$ \calK $ function. Then, the time-varying functions $ \bm{\lambda}_{i} $ can be determined as a solution to the differential equation $\dot{\bm{\lambda}}_{i}(t) =-\alpha_{\lambda}(\bm{\lambda}_{i}(t))$ with $\bm{\lambda}_{i}(T_{i})=0$ as boundary condition, and we obtain 
\begin{align*}
	\bm{\lambda}_{i}(t) = R^{2} u_{\text{max}}^{2} (t-T_{i})^{2} \qquad \forall t\leq T_{i}.
\end{align*}
By Theorem~\ref{thm:time-varying CBF without input constraints}, $ B_{\bm{\lambda}_{i}}(t,x) \coloneq b\big(x-[q_{i}^{T},0]^{T}\big)+\bm{\lambda}_{i}(t) $ is a uniformly time-varying CBF, and we piecewise define $B_{\bm{\lambda}}(t,x)\coloneq B_{\bm{\lambda}_{i}}(t,x)\big|_{i:\, T_{i-1}\leq t < T_{i})}$; the forward invariance of its zero super-level set follows from Corollary~\ref{corollary:cbf invariance}. Note that the square root in $ \alpha $ allows $ \bm{\lambda}_{i} $ to reach zero within finite time. This is in contrast to a linear $ \alpha $ that allows at most for an exponential decay rate. The simulation results are depicted in Figure~\ref{fig:sim_omni}. Time-varying function~$ \bm{\lambda} $, $ b $ and $ B_{\bm{\lambda}(\cdot)} $ are shown in Figure~\ref{fig:sim_omni}b. As $ B_{\bm{\lambda}(\cdot)} $ is positive for all times, all waypoints are reached within the time bounds. 
The resulting state trajectory can be seen from Figure~\ref{fig:sim_omni}a.

\subsection{Second Order System: Pendulum}

As an example for a second order system, consider the mechanical pendulum as in Sontag \cite[Example 5.7.5]{Sontag1998}, however slightly modified by an additional destabilizing momentum. Its dynamics are given as
\begin{align*}
	\dot{x}_{1} &= x_{2} \\
	\dot{x}_{2} &= -\frac{g}{l} \sin{(x_{1})} + d_{m}(x_{2}) + u
\end{align*} 
where $ x_{1} = \theta $ and $ x_{2} = \dot{\theta} $ are the excitation angle and velocity, respectively, $ u\in\calU\subseteq\bbR $ is the control input, $ d_{m}(x_{2}) = 5 l \, x_{2} $ is some destabilizing momentum, and we choose $ g=9.81 $ and $ l=1 $; the state vector is  $ x = [x_{1}, \, x_{2}]^{T} \in \bbR^{2} $. The pendulum in \cite{Sontag1998} comes along with a CLF, namely $ V(x) = 2 x_{1}^{2} + x_{2}^{2} + 2 x_{1} x_{2} $, which we shall use for the construction of a uniformly time-varying CBF. It can be shown that for some $ u\in\calU = [-20,20] $ and all $ x $ with $ V(x) \leq 2 $, we have
\begin{align*}
	&-\frac{\partial V}{\partial x}(x) \, f(x,u) \geq \max\{2V(x), 2x_{2}^{2}\} \\
	&\quad \geq \gamma(V\!(x)) \!=\!
	\begin{cases}
		V(x) & \text{for } V(x) \!<\! 0.03, \\
		0.03 \!+\! 2 (V\!(x)\!-\!0.03) & \text{otherwise},
	\end{cases}
\end{align*}
where $ \gamma: \bbR_{\geq0}\rightarrow\bbR_{\geq0} $ is a class~$ \calK $ function. By Theorem~\ref{thm:clf and cbf}, we know that $ b(x) := -V(x) $ is $ \Lambda $-shiftable with $ \Lambda = 2 $; its level sets are depicted in Figure~\ref{fig:sim_pendulum}a. The extended class~$ \calK_{e} $ function $ \alpha $ corresponding to $ b $ is constructed from $ \gamma $ as in the proof of Theorem~\ref{thm:clf and cbf}. As $ \gamma $ is convex, choose class~$ \calK $ function $ \alpha_{\lambda} $ as $ \alpha_{\lambda}(\xi) = -\alpha_{\lambda}(-\xi) = \gamma(\xi) $ for all $ \xi\in[0,\Lambda] $.
Based on this, we construct $ \bm{\lambda} $ as before such that~\eqref{eq:lambda_dot condition} holds for all $ t\geq0 $; the resulting trajectory $ \bm{\lambda} $ is depicted in Figure~\ref{fig:sim_pendulum}b. In particular for $ t\in[t_{1},t_{2}] $, we construct $ \bm{\lambda} $ such that~\eqref{eq:lambda_dot condition} holds with equality, which requires to solve a differential equation. This yields the fastest possible decreasing $ \bm{\lambda} $ such that~\eqref{eq:lambda_dot condition} still holds. On the other intervals, we construct $ \bm{\lambda} $ as linear functions, which allow to easily verify~\eqref{eq:lambda_dot condition}. By Theorem~\ref{thm:time-varying CBF without input constraints}, we conclude that $ B_{\bm{\lambda}(\cdot)}(t,x) := b(x) + \bm{\lambda}(t) $ is a CBF, and the forward invariance of its zero super-level set follows again from Corollary~\ref{corollary:cbf invariance}. As seen from Figure~\ref{fig:sim_pendulum}, the angle of excitation can be effectively controlled via the choice of $ \bm{\lambda} $, which varies the level set of $ b $ that is rendered forward invariant. Figure~\ref{fig:sim_pendulum}a clearly illustrates that the ability to constrain the angle of excitation by shifting function~$b$ through~$\bm{\lambda}$ strongly depends on the geometry of its level sets. While the analytical construction of a shiftable CBF and its associated extended class~$\mathcal{K}_{e}$ function provides conceptual insight, simultaneously accounting for both the state constraints and the dynamics may be nontrivial. The approach presented in Section~\ref{subsec:example time-varying obstacle} addresses this challenge by employing a numerical CBF construction.

\subsection{Linear System: 3d-Double Integrator Subject to Gravity}

Consider a double integrator in three dimensions subject to gravity, which can be viewed as the simplified model of a quadcopter for position control. Its dynamics are given as
\begin{align*}
	\dot{p} &= v\\
	\dot{v} &= \tfrac{1}{m} \bm{I}_{3} u + \left[\begin{smallmatrix}
		0 \\ 0 \\ -\tfrac{g}{m}
	\end{smallmatrix}\right]
\end{align*}
with position and velocity $ p,v\in\bbR^{3} $, and input $ u\in\bbR^{3} $; the mass and gravity constant are $ m=1.3 $ and $ g=9.81 $, respectively. The dynamics can be equivalently written as 
$ \dot{x} = \left[\begin{smallmatrix}
	\bm{0} & \bm{I}_{3} \\ \bm{0} & \bm{0}
\end{smallmatrix}\right] x  + \left[\begin{smallmatrix}
\bm{0} \\ \tfrac{1}{m} \bm{I}_{3}
\end{smallmatrix}\right] \Delta u $ 
with $ x = [p^{T}, v^{T}]^{T} $ and $ \Delta u \coloneq u - [0,0,g]^{T} $. Similarly to Section~\ref{subsec:sim_omni}, the objective is to reach the origin as quickly as possible by shrinking the zero super-level set of a shiftable CBF. Unlike the previous example, the focus is on comparing the conservativeness of analytically and numerically computed extended class~$\mathcal{K}_{e}$ functions and evaluating its effect on the time required to reach the destination. By following Example~\ref{example:lqr}, we first design a CLF and its associated class~$ \calK $ function based on an LQR controller with $ Q=\bm{I}_{6} $, $ R = 6\bm{I}_{3} $, from which a $ \Lambda $-shiftable CBF directly follows via Theorem~\ref{thm:clf and cbf}. For $ \Lambda=100 $, the satisfaction of input constraint $ ||\Delta u||_{\infty} \leq 6.5 $ can be guaranteed for the chosen $ R $. The resulting LQR-based CBF takes the form $ b(x) = ||x||_{P}^{2} $ with the associated extended class~$ \calK_{e} $ function $ \alpha(b) \coloneq c_{\alpha} b = 0.1 b $. The analytic computation of $ \alpha $, however, is generally conservative. The least conservative extended class~$ \calK_{e} $ function is numerically obtained for $ c_{\alpha} = 0.7 $. We construct a uniformly time-varying CBF with $ \bm{\lambda} $ obtained by solving the differential equation $ \dot{\bm{\lambda}}(t) = \alpha(-\bm{\lambda}(t)) $, which allows the system to reach the neighborhood of a waypoint within the shortest time that is admitted by the respective $ \alpha $ function. The simulation results are depicted in Figure~\ref{fig:sim_quadcop}. Clearly, the least conservative $ \alpha $ allows the system to reach the destination significantly faster than the analytic one. 

\subsection{Non-Holonomic System: Unicycle}
\label{subsec:example nonholonomic system}

\begin{figure}[t]
	\centering
	\begin{subfigure}[b]{0.52\linewidth}
		\centering
		\includegraphics[width=1.0\linewidth]{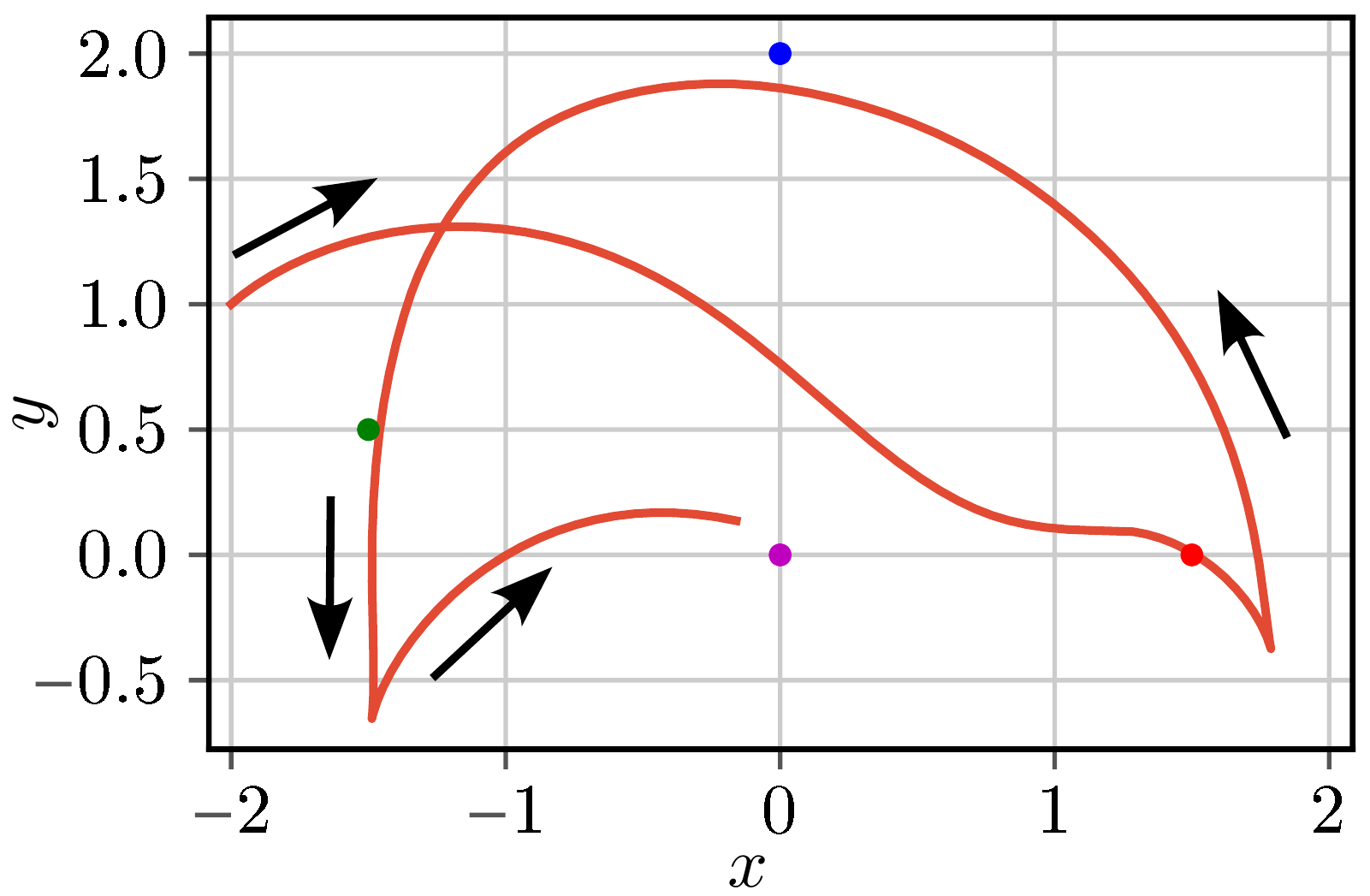}
		\caption{State trajectory.}
		\label{sub_fig:sim_nint_traj}
	\end{subfigure}
	\hfill
	\begin{subfigure}[b]{0.44\linewidth}
		\centering
		\includegraphics[width=1.0\linewidth]{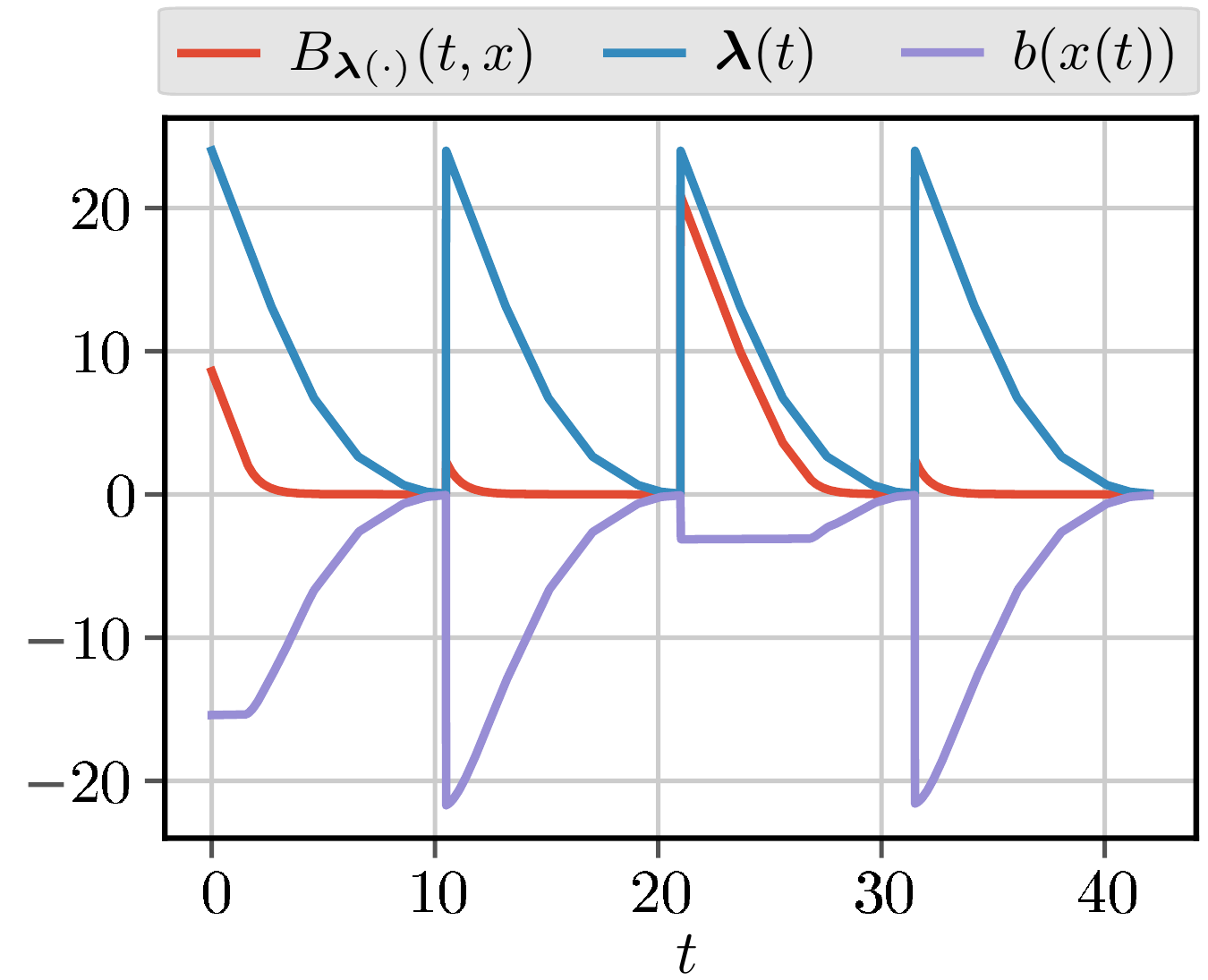}
		\caption{Value functions.}
		\label{sub_fig:sim_nint_cbf_traj}
	\end{subfigure}
	\caption{Invariant set-based waypoint tracking for a unicycle.}
	\label{fig:sim_nint}
\end{figure}

\begin{figure*}[t]
	\centering
	\begin{minipage}[t]{0.55\textwidth}
		\begin{subfigure}[b]{0.6\linewidth}
			\centering
			\includegraphics[width=1.0\linewidth]{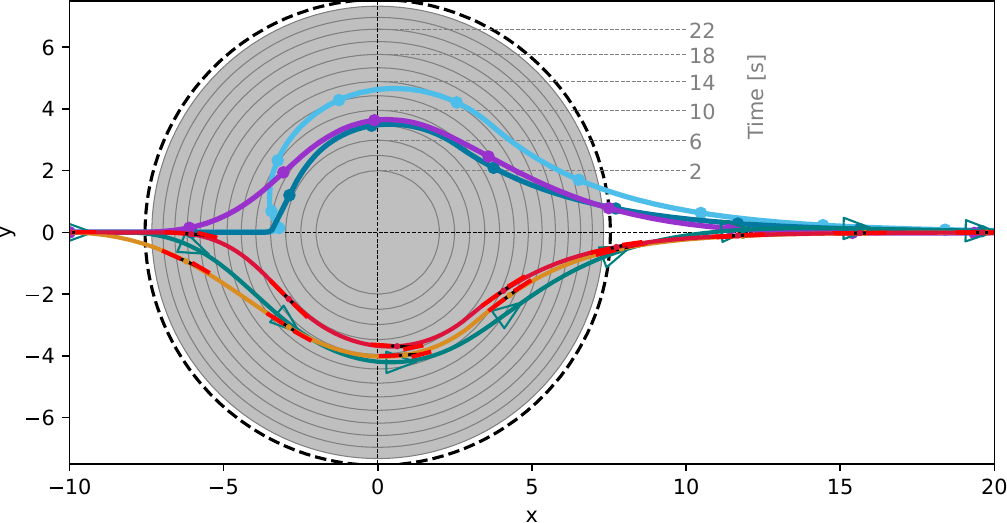}
			\caption{Trajectories and obstacle over time. Every 2s a marker is depicted.}
			\label{sub_fig:sim_obst_traj}
		\end{subfigure}
		\hfill
		\begin{subfigure}[b]{0.35\linewidth}
			\centering
			\includegraphics[width=\linewidth]{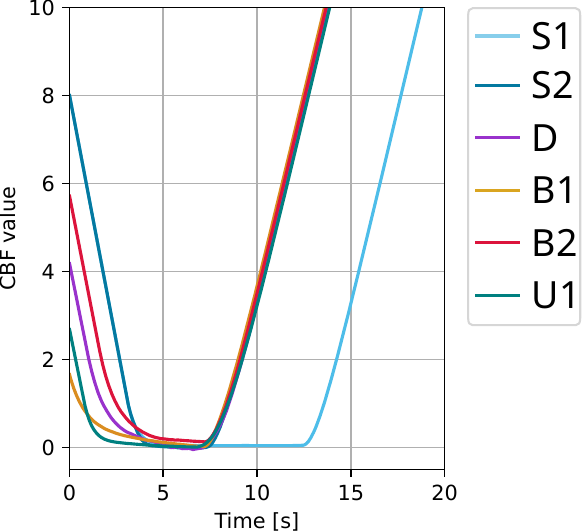}
			\caption{Value functions over time.}
			\label{sub_fig:sim_obst_cbf_traj}
		\end{subfigure}
		\caption{Simulation results for the avoidance of a time-varying obstacle.}
		\label{fig:sim_obstacle}
	\end{minipage}
	\hfill
	\begin{minipage}[t]{0.4\textwidth}
		\begin{subfigure}[b]{0.48\linewidth}
			\centering
			\includegraphics[width=1.0\linewidth]{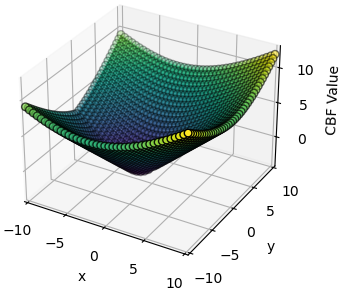}
			\caption{Shiftable CBF for unicycle at $ \psi=0 $.}
			\label{sub_fig:sim_obst_u_cbf}
		\end{subfigure}
		\hfill
		\begin{subfigure}[b]{0.48\linewidth}
			\centering
			\includegraphics[width=1.0\linewidth]{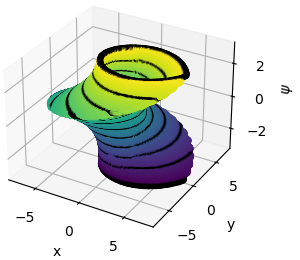}
			\caption{Zero super-level set of the shiftable CBF.}
			\label{sub_fig:sim_obst_u_zero_level}
		\end{subfigure}
		\caption{Nonsmooth shiftable CBF in the Dini sense for the unicycle.}
		\label{fig:sim_obstacle_cbf_u}
	\end{minipage}
\end{figure*}

Our next example addresses a non-holonomic system, namely the unicyle. It aims, as the examples before, on the tracking of a sequence of waypoints within prescribed time through a uniformly time-varying CBF, yet with the complication of the non-holonomic constraint. The unicycle dynamics are given as
\begin{align}
	\label{eq:unicycle dynamics}
	\begin{split}
		\dot{x} &= v\cos(\theta), \\
		\dot{y} &= v\sin(\theta), \\
		\dot{\theta} &= \omega.
	\end{split}
\end{align}
We construct a non-smooth shiftable CBF in the Dini sense based on the CLF in the Dini sense given in~\cite{Clarke2011} for the non-holonomic integrator, which is 
\begin{align*}
	V(\tilde{x}) = \left(\sqrt{\tilde{x}_{1}^{2} + \tilde{x}_{2}^{2}} - |\tilde{x}_{3}|\right)^{2} + \tilde{x}_{3}^{2},
\end{align*}
where $ \tilde{x} $ denotes the state of the non-holonomic integrator and it holds $ \dot{V}(\tilde{x}) \leq -W(\tilde{x}) $ with $ W(\tilde{x}) \coloneq 2\max\{|\sigma - |\tilde{x}_{3}||,\sigma|\sigma\text{sgn}(\tilde{x}_{3}) - 2x_{3}|\} $ and $ \sigma\coloneq\sqrt{\tilde{x}_{1}^{2}+\tilde{x}_{2}^{2}} $. The extended class~$ \calK $ function~$ \gamma $ corresponding to the CLF (see Definition~\ref{def:clf_dini}) is determined such that 
\begin{align*}
	\gamma(\xi) \leq \min_{x\!: V(x)=\xi} W(x).
\end{align*}
This is the case for 
\begin{align*}
	\gamma(\xi) = \left\{\begin{smallmatrix}
		\tfrac{1}{\sqrt{3}} \, \xi \hfill & \text{if } |\xi|<3, \hfill \\
		1.3\, \text{sgn}(\xi) \sqrt{|\xi|} \hfill & \text{otherwise.} \hfill
	\end{smallmatrix}\right.
\end{align*}
Finally, the shiftable CBF is obtained via Theorem~\ref{thm:clf and cbf} and the diffeomorphism relating the non-holonomic integrator to the unicycle dynamics~\cite{DeVon2007}. From here on, the controller design is analogous to Section~\ref{subsec:sim_omni}. The simulation results are shown in Figure~\ref{fig:sim_nint}. Note, that in particular the resulting state trajectory depicted in Figure~\ref{sub_fig:sim_nint_traj} shows a behavior that rather resembles that of a prediction-based controller. This illustrates that CBFs, and CLFs and shiftable CBFs in particular, provide a characterization of the system's dynamic capabilities with respect to a constraint.

\subsection{Avoidance of a Uniformly Time-Varying Obstacle}
\label{subsec:example time-varying obstacle}

At last, we consider the avoidance of an obstacle, whose size is increasing over time, for various dynamic systems tracking a reference. Specifically, we consider:  
(1)~two single integrators $ \dot{x} = u $, where the second one can only move forward into the strictly positive $ x $-direction ($ \calU_{\text{S1}} = [-2,2]^{2} $ and $ \calU_{\text{S2}} = [1,2]\times[-2,2] $); (2)~a double integrator $ \ddot{x} = u $ with $ u\in\calU_{\text{D}} = [-1,1]^{2} $; (3)~two bicycle models~\cite{Wang2001} with dynamics $ \dot{x} = v \cos(\psi+\beta(\zeta)) $, $ \dot{y} = v \sin(\psi + \beta(\zeta)) $, $ \dot{\psi} = \frac{v}{L} \cos(\beta(\zeta)) \tan(\zeta) $, with $ \beta(\zeta) = \arctan(\frac{1}{2} \tan(\zeta)) $, and velocity $ v\in[1,2] $ and steering angle $ \zeta $ as inputs; the steering angle of the first (less agile) system is constrained by $ |\zeta| \leq \frac{20}{180}\pi $ and the one of the second (more agile) system by $ |\zeta|\leq\frac{40}{180}\pi $; (4) a unicycle with dynamics~\eqref{eq:unicycle dynamics} and input constraints $ v\in[1,2] $, $ \omega\in[-0.9,0.9] $. 

For each system, shiftable CBFs and the corresponding extended class~$\mathcal{K}_e$ functions are numerically computed via the reachability-based CBF synthesis method in~\cite{Wiltz2025b}, using the same design parameters as in~\cite[Table~1]{Wiltz2025b}. The uniformly time-varying CBFs and their corresponding time-varying functions~$\bm{\lambda}$ are derived for the considered time-varying constraint as outlined in Section~\ref{sub_sec:application to time-varying constraints}. The simulations are depicted in~Figure~\ref{fig:sim_obstacle}. As it can be seen from Figure~\ref{sub_fig:sim_obst_traj}, the single-integrator S1, that can move freely in all directions, approaches the time-varying obstacle until it reaches its boundary, subsequently moves partially backwards while circumventing it. In contrast, the remaining systems are less agile due to their dynamics and input constraints, requiring them to anticipate the obstacle motion and initiate avoidance maneuvers earlier. The underlying shiftable CBFs, exemplarily depicted in Figure~\ref{fig:sim_obstacle_cbf_u} for the input constrained  unicycle, encode exactly the information required for this anticipatory behavior. Note that the numerically computed shiftable CBF for the unicycle is clearly nonsmooth.

%% file: 20-conclusion.tex
\section{Conclusion}
\label{sec:conclusion}

We presented a systematic framework for designing uniformly time-varying CBFs by decomposing the problem into the design of a time-invariant and a time-varying component. We characterized the relevant subclass of time-invariant CBFs, termed shiftable CBFs, and derived conditions on time-varying functions $ \bm{\lambda} $ preserving the CBF property of the time-invariant CBF when added. This enables handling diverse time-variations in the state constraints without redesigning the time-invariant shiftable CBF. Our analysis revealed a direct link between the extended class~$ \calK_{e} $ function $ \alpha $ associated with a shiftable CBF and the rate of change of~$ \bm{\lambda} $, highlighting the importance of a sophisticated choice of~$ \alpha $. We further established forward invariance results for the case when the uniformly time-varying value function is not a CBF. Finally, we pointed out how existing CBF and CLF construction methods can be employed in designing shiftable CBFs, and demonstrated the effectiveness of our framework through numerical examples. Identifying further classes of time-varying CBFs that support such a decoupled design remains an open problem for future work. Furthermore, extending results on control sharing and compatible CBFs to the time-varying setting would broaden the applicability of uniformly time-varying CBFs to problems involving multiple time-varying constraints.